\documentclass[12pt,floats,aps]{article}
\usepackage{epsfig}
\usepackage{a4wide}

\newcommand{\beq}{\begin{equation}}
\newcommand{\eeq}{\end{equation}}
\newcommand{\bea}{\begin{eqnarray}}
\newcommand{\eea}{\end{eqnarray}}

\newcommand{\Tr}{{\,\rm Tr}\:}

\newcommand{\moy}[1]{\left<{#1}\right>}

\newcommand{\td}{\tilde}

\begin{document}

\hfill{SPhT: T04/128}
\vspace{0.5cm}

\begin{center}
{\bf Mod\`eles \`a 1 et 2 matrices et fonction g\'en\'eratrice de cylindres \`a bords bicolores.}
\vspace{0.8cm}

N. Orantin\footnote{or{an}tin@spht.sacl{ay.c{e}a.}fr}

{\emph
Service de Physique th\'eorique, CEA/Saclay

Orme des Merisiers F-91191 Gif-sur-Yvette Cedex, France}
\end{center}
\vspace{1.5cm}

\begin{center}
{\bf Abstract:}
\end{center}

Je vais traiter dans ce rapport des mod\`{e}les \`{a} une et deux matrices
comme outil pour \'{e}tudier les th\'{e}ories conformes avec bords. J'introduirai dans un premier temps certains r\'{e}sultats
bien connus que le lecteur pourra trouver en grande partie dans les ouvrages de r\'{e}f\'{e}rences que sont
\cite{Mehta}, \cite{kunz} et \cite{courseynard} pour ce qui est des matrices al\'{e}atoires et \cite{Farkas} pour la g\'{e}om\'{e}trie alg\'{e}brique.
Pour ce faire, apr\`{e}s une rapide pr\'{e}sentation historique des matrices al\'{e}atoires et de leurs applications,
je m'int\'{e}resserai tour \`{a} tour aux mod\`{e}les \`{a} une et deux matrices, pr\'{e}sentant leur formalisme,
expression en termes de diagrammes de Feynmann et surfaces al\'{e}atoires ainsi que leur lien avec les th\'{e}ories conforme.
Je conclurai sur le mod\`{e}le \`{a} deux matrices en expliquant la m\'{e}thode des \'{e}quations de boucles.
Je m'attacherai dans un second temps \`{a} pr\'{e}senter un travail de recherche personnel consistant en le calcul
de la fonction de corr\'{e}lation g\'{e}n\'{e}ratrice des surfaces al\'{e}atoires ayant la forme d'un cylindre
aux bords bicolores.

\section{Introduction de la notion de matrices al\'{e}atoires et de
leurs applications.}

Parmi les nombreuses applications des matrices al\'{e}atoires en physique que j'aborderai plus bas,
je me suis particuli\`{e}rement int\'{e}ress\'{e} \`{a} leur interpr\'{e}tation en termes de surfaces al\'{e}atoires.
En effet, les int\'{e}grales de matrices peuvent \^{e}tre vues comme les fonctions g\'{e}n\'{e}ratrices de
surfaces discretis\'{e}es avec ou sans bords. Les op\'{e}rateurs vivant sur ces surfaces sont d\'{e}crit
par les th\'{e}ories conformes dans la limite o\`{u} les surfaces discretis\'{e}es deviennent continues.
Les exposants critiques de tels op\'{e}rateurs sont bien connus lorsque ceux ci vivent \`{a} l'int\'{e}rieur
de la surface (\cite{KPZ}). Les op\'{e}rateurs agissant sur les bords sont quant \`{a} eux bien moins connus
et les mod\`{e}les de matrices semblent \^{e}tre un bon outil pour d\'{e}terminer leurs exposants critiques.
Les fonctions g\'{e}n\'{e}ratrices des surfaces avec deux et quatre op\'{e}rateurs sur un m\^{e}me bord
ayant d\'{e}j\`{a} \'{e}t\'{e} d\'{e}termin\'{e}es (\cite{eynardchain}, \cite{eynm2m}), ces notes ont pour but
de pr\'{e}senter le calcul de la fonction g\'{e}n\'{e}ratrice des surfaces avec deux bords portant chacun
deux op\'{e}rateurs.

\subsection{Historique.}
Le terme al\'{e}atoire associ\'{e} aux matrices dont je vais
introduire l'\'{e}tude dans ce rapport semblerait induire que ces
derni\`{e}res sont apparues comme un outil pour la physique
statistique. Cependant, il en a \'{e}t\'{e} tout autrement.

En effet, elles ont vu le jour dans le cadre de la physique
atomique et plus pr\'{e}cis\'{e}ment dans le cadre de l'\'{e}tude de
noyaux lourds par Wigner en 1951 (\cite{Mehta}). Le spectre en \'{e}nergie de ces
derniers \'{e}tant trop dense pour pouvoir tenir compte
s\'{e}par\'{e}ment de chaque niveau, Wigner a consid\'{e}r\'{e} le
probl\`{e}me d'un point de vue statistique. Il a observ\'{e} que
la distribution moyenne des niveaux d'\'{e}nergie, des fonctions
de corr\'{e}lation ainsi que d'autres param\`{e}tres tels que
l'espacement moyen entre les niveaux d'\'{e}nergie,
correspondaient au spectre d'une matrice al\'{e}atoire suivant une
loi de probabilit\'{e} gaussienne prise dans un ensemble
d\'{e}pendant du probl\`{e}me consid\'{e}r\'{e} (je reviendrai
plus loin sur la correspondance entre ensemble de matrices et
caract\'{e}ristiques du probl\`{e}me consid\'{e}r\'{e}). La
surprenante fid\'{e}lit\'{e} des r\'{e}sultats exp\'{e}rimentaux
avec cette description en termes de matrices imposa pour la
premi\`{e}re fois les matrices al\'{e}atoires comme un puissant
outil math\'{e}matique de description de syst\`{e}mes physiques
complexes.

D\`{e}s lors, les matrices al\'{e}atoires ont connu un succ\`{e}s
sans cesses grandissant en apparaissant dans de nombreux domaines
de la physique pourtant \'{e}loign\'{e}s les uns des autres. Avant
d'aller plus avant dans une pr\'{e}sentation technique de cet
outil, je vais bri\`{e}vement introduire certains de ses principaux
domaines d'application ainsi que l'une de ses principales
caract\'{e}ristiques: le principe d'universalit\'{e}.

\subsection{Le principe d'universalit\'{e}.}
On peut consid\'{e}rer la notion de matrice al\'{e}atoire comme
une g\'{e}n\'{e}ralisation des nombres al\'{e}atoires. On est
alors en droit de se demander s'il existe un
\'{e}quivalent au th\'{e}or\`{e}me centrale-limite dans le cas des
matrices. Ce comportement particulier aux grands nombres peut
\^{e}tre observ\'{e} lorsque l'on fait tendre la taille des
matrices vers l'infini. Le spectre obtenu dans cette limite ne
d\'{e}pend plus de la loi de probabilit\'{e} initiale mais
simplement des sym\'{e}tries de l'ensemble de matrices
consid\'{e}r\'{e}.

Ce comportement n'est, pour l'instant, que conjectur\'{e}.
Cependant, les v\'{e}rifications exp\'{e}rimentales ainsi que des
d\'{e}monstrations dans des cas pr\'{e}cis permettent d'avoir
suffisamment confiance en cette propri\'{e}t\'{e} pour en faire
l'une des principales composantes du domaine. Elle
permet en effet de conna\^\i tre le comportement de n'importe quel
ensemble de matrices, \`{a} grand nombre de variables, en
utilisant une loi de probabilit\'{e} gaussienne.

 Il est alors important de classer les diff\'{e}rents ensembles de matrices
selon leurs propri\'{e}t\'{e}s de sym\'{e}tries.
On rencontre trois cat\'{e}gories de tels ensembles selon le point
de vue que l'on adopte, mais je n'en pr\'{e}senterai ici que deux, la troisi\`{e}me
correspondant aux matrices de transfert. D'une part, on peut consid\'{e}rer le
hamiltonien comme une matrice al\'{e}atoire suivant une loi de
probabilit\'{e} gaussienne. On devra alors consid\'{e}rer les
ensembles suivants:
\begin{itemize}
    \item Lorsqu'il n'y a aucune
    sym\'{e}trie particuli\`{e}re, on consid\`{e}re l'ensemble des
    matrices hermitiennes, appel\'{e} Gaussian Unitary Ensemble
    (GUE), car il est invariant par les transformations unitaires;
    \item Lorsque l'on ajoute une invariance du syst\`{e}me par
    renversement du temps, il convient alors de consid\'{e}rer
    l'ensemble des matrices r\'{e}elles sym\'{e}triques,
    appel\'{e} Gaussian Orthogonal Ensemble (GOE);
    \item Si, de plus, le syst\`{e}me a un spin total demi-entier
    et que la sym\'{e}trie de rotation est bris\'{e}e, on
    utilisera l'ensemble des matrices quaternioniques self-duales
    r\'{e}elles, appel\'{e} Gaussian Symplectic Ensemble (GSE).
\end{itemize}

D'autre part, on peut consid\'{e}rer l'op\'{e}rateur
d'\'{e}volution comme une matrice al\'{e}atoire suivant une loi
de probabilit\'{e} constante, ce qui fait appara\^\i tre les ensembles
suivants:
\begin{itemize}
    \item Sans sym\'{e}trie particuli\`{e}re, on aura l'ensemble
    des matrices unitaire ou Circular Unitary Ensemble (CUE);
    \item Avec renversement du temps, on prendra les matrices
    orthogonales, soit le Circular Orthogonal Ensemble (COE);
    \item Enfin, on peut consid\'{e}rer l'ensemble des matrices
    symplectiques, le Circular Symplectic Ensemble (CSE).
\end{itemize}

\subsection{Quelques applications.}
L'id\'{e}e la plus naturelle d'utilisation des matrices
al\'{e}atoires consiste simplement \`{a} consid\'{e}rer un
syst\`{e}me d\'{e}sordonn\'{e} (\cite{Guhr}) o\`{u} le hamiltonien est
lui-m\^{e}me une matrice al\'{e}atoire. Selon les sym\'{e}tries pr\'{e}c\'{e}demment
\'{e}num\'{e}r\'{e}es, on aura le hamiltonien dans les ensembles
GUE, GOE ou GSE.

Une seconde application intervient lors de l'\'{e}tude d'un
probl\`{e}me tout aussi naturel: le chaos quantique (\cite{Guhr}). Si
l'\'{e}tude des syst\`{e}mes chaotiques est bien connue en
m\'{e}canique classique, sa g\'{e}n\'{e}ralisation en
m\'{e}canique quantique est tr\`{e}s probl\'{e}matique.
L\`{a} encore, les matrices al\'{e}atoires apparaissent
indirectement. En effet, le hamiltonien du syst\`{e}me
consid\'{e}r\'{e} n'est pas \`{a} proprement parler al\'{e}atoire.
Cependant, les observations ont montr\'{e} que, si le probl\`{e}me
classique associ\'{e} est chaotique, son spectre est le m\^{e}me
que celui d'une matrice al\'{e}atoire appartenant aux ensembles
GUE, GOE ou GSE. On peut \'{e}galement consid\'{e}rer
l'op\'{e}rateur d'\'{e}volution qui, lui, appartiendra \`{a} l'un
des ensembles CUE, COE ou CSE. Notons que cette relation entre
chaos quantique et matrices al\'{e}atoires n'est pas
d\'{e}montr\'{e}e mais simplement observ\'{e}e exp\'{e}rimentalement et
conjectur\'{e}e th\'{e}oriquement.

Les matrices al\'{e}atoires apparaissent \'{e}galement en
chromodynamique quantique (QCD). Cette th\'{e}orie suppose
l'existence de particules fondamentales constitutives des hadron,
les quarks et les gluons, qui caract\'{e}risent l'\'{e}change de la charge de couleur (de dimension N=3)
entre les quarks. Cet \'{e}change se traduit par une matrice $3 \times 3$. De mani\`{e}re analogue \`{a}
la m\'{e}canique quantique o\`{u} l'on doit int\'{e}grer sur tous les \'{e}tats  possibles, il est ici n\'{e}cessaire
de consid\'{e}rer un ensemble de matrices pond\'{e}r\'{e} par une loi de probabilit\'{e}. L'utilisation du
mod\`{e}le de matrices al\'{e}atoires a permis d'obtenir des
pr\'{e}dictions non perturbatives. Pour ce faire,il a \'{e}t\'{e}
propos\'{e} par 't Hooft (\cite{thoft})de faire un d\'{e}veloppement en
puissance de 1/N quand le nombre de couleur N est artificiellement
amen\'{e} \`{a} tendre vers $+ \infty$. Ce d\'{e}veloppement en
puissances de 1/N est en fait un d\'{e}veloppement topologique sur
lequel nous reviendrons plus loin. On peut \'{e}galement profiter du principe d'universalit\'{e} en faisant tendre
le nombre de degr\'{e}s de libert\'{e}s du syst\`{e}me vers l'infini (\cite{Verbaarshot}). Enfin, cet outil
math\'{e}matique permet de faire un lien entre la QCD et la
th\'{e}orie des cordes (\cite{ZJDFG}, \cite{RMTGQ}, \cite{GinspargGQ2D}, \cite{grossGQ2D}) , sujet que nous allons aborder dans le
prochain paragraphe.

Les matrices al\'{e}atoires et la th\'{e}orie des cordes se sont
rencontr\'{e}es \`{a} plusieurs reprises (\cite{ZJDFG}, \cite{DVV}, \cite{Dijgrafvafa}). Je ne vais \'{e}voquer
ici qu'un aspect de cette relation. Comme nous le verrons plus
loin, les matrices al\'{e}atoires permettent, gr\^{a}ce au
d\'{e}veloppement topologique mis en place pour la QCD, de
d\'{e}crire et de sommer des surfaces al\'{e}atoires. Or, les
cordes \'{e}voluant dans le temps, forment justement de telles
surfaces (\cite{Bilal:1997fy}, \cite{Antoniadis:1999y}). Il est donc naturel d'utiliser un mod\`{e}le de matrices
pour traiter un tel probl\`{e}me. Bien que les mod\`{e}les de
matrices permettent d'avoir une th\'{e}orie de la sommation des
surfaces, ils sont limit\'{e}s \`{a} l'\'{e}tude de la th\'{e}orie
des cordes en dimension $D \leq 1$, ce qui emp\'{e}che d'atteindre
le cas physique $D = 10,26$. Cependant, les mod\`{e}les de matrices
restent int\'{e}ressants que ce soit comme toy-model pour $D \leq 1$
ou gr\^{a}ce aux autres liens qu'ils entretiennent avec la
th\'{e}orie des cordes et les r\'{e}centes tentatives d'unifications
comme la th\'{e}orie M (\cite{DVV}, \cite{BFSS}, \cite{IKKT}).

Au cours de cet expos\'{e}, je reviendrai plus longuement sur deux autres importantes applications:
l'\'{e}tude des surfaces al\'{e}atoires et celle des th\'{e}ories conformes.

\section{Mod\`{e}le \`{a} une matrice.}

\subsection{Introduction des notations.}
En premier lieu, introduisons les notations utilis\'{e}es tout au
long de ces notes ainsi que certaines consid\'{e}rations
g\'{e}n\'{e}rales.

Soit un ensemble E de matrices de taille $N \times N$ sur lequel
on d\'{e}fini une loi de probabilit\'{e} P(M) comme un poids de
Boltzmann:
\begin{equation}
P(M) = \frac{1}{Z} e^{-\frac{N}{T} Tr V(M)} dM
\end{equation}

o\`{u} le potentiel V(M), la mesure dM (dans le cas o\`{u} E est l'ensemble des matrices hermitiennes) et la fonction de partition
Z sont d\'{e}finis par:
\begin{equation}
\begin{array}{c}
dM = \prod_{i=1}^{N} \, dM_{ii} \,\, \prod_{i<j} \, dRe M_{ij} \,\, dIm M_{ij}\\
Z = \int_E dM e^{-\frac{N}{T} Tr V(M)} = e^{-\frac{N^2}{T^2} F}\\
V(x) = V_0 + \sum_{k=1}^{d+1} \frac{t_k}{k} x^k\\
\end{array}
\end{equation}

\textbf{Note :} le cas gaussien \'{e}voqu\'{e} pr\'{e}c\'{e}demment
correspond au potentiel $ V(x) = \frac{t_2}{2} x^2 $.
\vspace{0.5cm}

 Nous avons introduit ici l'\'{e}nergie libre F qui
est une fonction fondamentale dans l'\'{e}tude des matrices
al\'{e}atoires. On montrera ainsi facilement que :
\begin{equation}
\begin{array}{c}
k \frac{\partial F}{\partial t_k} = \frac{1}{N} \left\langle Tr
M^k \right\rangle\\
\frac{\partial^2 F}{\partial t_k \partial t_j} = - \frac{1}{k j}
\left\langle Tr M^k \, Tr M^j \right\rangle_c\\
.\\
.\\
.\\
k_1 ... k_n \frac{\partial^n F}{\partial t_{k_1} ... \partial
t_{k_n}} = (-1)^{n-1} N^{n-2}\left\langle Tr M^{k_1} ... Tr M^{k_n} \right\rangle_c\\
\end{array}
\end{equation}
o\`{u} l'on a not\'{e} $\langle \prod x_i \rangle = \langle
\prod x_i \rangle_c + \sum_i \langle x_i \rangle \langle \prod_{j \neq i} x_j \rangle_c + 
+ \sum_{i,j \neq i} \langle x_i x_j \rangle_c \langle \prod_{k \neq i,j} x_k \rangle_c + ... + \prod \langle x_i \rangle$.

Notons que les d\'{e}riv\'{e}es secondes de l'\'{e}nergie libre
par rapport aux $t_i$
correspondent \`{a} des fonctions de corr\'{e}lation \`{a} deux points et sont en
fait des fonctions universelles. De mani\`{e}re plus
g\'{e}n\'{e}rale, toute d\'{e}riv\'{e}e seconde de F est une telle
fonction universelle (\cite{courseynard}, \cite{AmbjAk}) et ce sont ces derni\`{e}res qui
sont porteuses du principe d'universalit\'{e}.

\subsection{Expression en termes de diagrammes de Feynman et de surfaces al\'{e}atoires.}
Comme il l'a \'{e}t\'{e} dit plus t\^{o}t, l'une des principales
applications des matrices al\'{e}atoires se rapporte \`{a} leur
interpr\'{e}tation en termes de diagrammes de Feynman (\cite{Matrixsurf}, \cite{RMTGQ}, \cite{BIPZ}).
\vspace{0.5cm}

Consid\'{e}rons l'ensemble E des matrices hermitiennes M. On
aimerait calculer la fonction de partition Z d\'{e}finie plus
haut. Afin d'obtenir une interpr\'{e}tation des mod\`{e}les de matrices en gravit\'{e} quantique,
 on va proc\'{e}der par un d\'{e}veloppement perturbatif
autour d'un potentiel gaussien. En effet, on sait effectuer ce
calcul dans le cas d'un potentiel gaussien et l'on pourra ainsi
obtenir le cas g\'{e}n\'{e}ral comme perturbation de
l'int\'{e}grale gaussienne. Notons l'analogie avec la physique des
particules o\`{u} le potentiel gaussien correspond \`{a} des
particules libres sans interaction, les int\'{e}grales
perturb\'{e}es \'{e}tant obtenues comme un d\'{e}veloppement en
diagrammes de Feynman. Dans le cas des matrices al\'{e}atoires, ce
d\'{e}veloppement analogue en diagrammes s'identifiera \`{a} un
d\'{e}veloppement topologique en termes de surfaces.
\vspace{0.5cm}

D\'{e}finissons donc $\delta V$ par :
\begin{equation}
V(M) = V_0 + \frac{t_2}{2} M^2 - \delta V(M)
\end{equation}

En consid\'{e}rant la partie quadratique au voisinage de son
minimum comme dominante et effectuant le d\'{e}veloppement \`{a}
$\delta V$ petit, on peut \'{e}crire le d\'{e}veloppement :
\begin{equation}
Z = e^{-\frac{N^2 V_0}{T}} \int_E dM e^{-\frac{N t_2}{2 T} Tr M^2}
( 1 + \frac{N}{T} Tr \delta V(M) + \frac{N^2}{2 T^2} [Tr \delta
V(M)]^2 + ... )
\end{equation}

Ce qui donne :
\bea
Z & = & e^{-\frac{N^2 V_0}{T}} \int_E dM e^{-\frac{N t_2}{2 T} Tr M^2} \left\langle e^{\frac{N}{T} Tr \delta
V} \right\rangle_0 \cr
& = & e^{-\frac{N^2 V_0}{T}} \int_E dM
e^{-\frac{N t_2}{2 T} Tr M^2} \left( 1 + \frac{N}{T} \left\langle Tr \delta
V \right\rangle_0 + \frac{N^2}{2 T^2} \left\langle (Tr \delta
V)^2\right\rangle_0 + ... \right)
\eea

o\`{u} l'indice 0 indique que l'on moyenne sur la distribution
gaussienne $\int dM e^{-\frac{N t_2}{2 T} Tr M^2}$.
\vspace{0.5cm}

D\'{e}s lors, le th\'{e}or\`{e}me de Wick permet de d\'{e}composer
les moyennes de produits de variables al\'{e}atoires en une somme
sur tous les appariements du produit des valeurs moyennes par
paires.

Par exemple:
\begin{equation}
\left\langle \phi_1 \phi_2 \phi_3 \phi_4\right\rangle_0 =
\left\langle \phi_1 \phi_2\right\rangle_0 \left\langle\phi_3
\phi_4\right\rangle_0 + \left\langle \phi_1 \phi_3\right\rangle_0
\left\langle\phi_2 \phi_4\right\rangle_0 + \left\langle \phi_1
\phi_4\right\rangle_0 \left\langle\phi_2 \phi_3\right\rangle_0
\end{equation}

C'est ce d\'{e}veloppement qui est \`{a} la base de la
repr\'{e}sentation diagrammatique de l'int\'{e}grale. Pour
repr\'{e}senter chaque terme de ce d\'{e}veloppement sous forme de
diagramme, nous allons introduire des r\`{e}gles de Feynman
expliquant comment les tracer et quel poids leur attribuer.

\subsubsection{Les r\`{e}gles de Feynman.} Les r\`{e}gles de Feynman
consistent \`{a} faire correspondre \`{a} chaque terme sous forme
d'une valeur moyenne, une repr\'{e}sentation diagrammatique
compos\'{e}e de vertex reli\'{e}s entre eux par des propagateurs.

La r\`{e}gle fondamentale consiste \`{a} d\'{e}finir le
propagateur correspondant \`{a} la valeur moyenne du produit d'une
paire de variables gaussiennes. Le propagateur est donn\'{e},
comme il est d'usage en th\'{e}orie des champs, par l'inverse de
la partie quadratique :

\begin{equation}
\begin{array}{r}
{\epsfxsize 3cm\epsffile{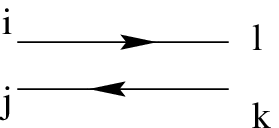}}
\end{array}= \left\langle M_i^j M_k^l \right\rangle_0 =
\frac{T}{N t_2} \delta_i^l \delta_j^k
\end{equation}

Notons que, contrairement \`{a} la th\'{e}orie classique des
champs, le propagateur n'est pas simplement une ligne non
orient\'{e}e, mais est compos\'{e} de doubles lignes
orient\'{e}es.
\vspace{0.5cm}

Les vertex, repr\'{e}sentant les termes de la forme $\left\langle
\frac{N t_k}{k T} Tr M^k \right\rangle_0$, sont quant \`{a} eux
issus de $\left\langle Tr \delta V \right\rangle_0.$ Chaque terme
de degr\'{e} k sera repr\'{e}sent\'{e} par un diagramme \`{a} k
lignes externes sous la forme de lignes doubles orient\'{e}es de
fa\c{c}on \`{a} pouvoir les connecter \`{a} des propagateurs. On a
alors, pour le terme de degr\'{e}s k du potentiel :
\begin{equation}
\begin{array}{r}
{\epsfxsize 4cm\epsffile{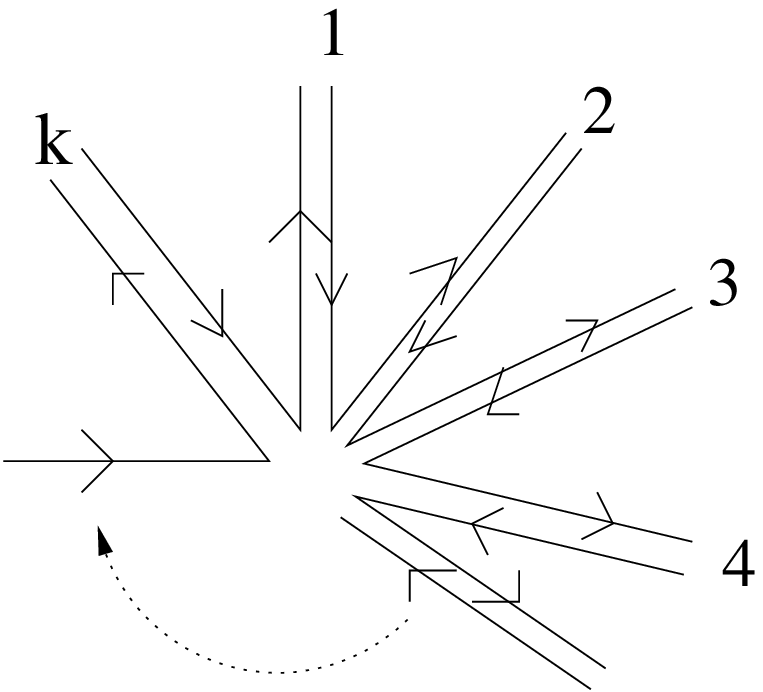}}
\end{array}= \frac{N t_k}{T k} \delta_{i_2}^{j_1}
\delta_{i_3}^{j_2} ... \delta_{i_k}^{j_{k-1}} \delta_{i_1}^{j_{k}}
\end{equation}

\subsubsection{D\'{e}termination du d\'{e}veloppement diagrammatique.}
Les r\`{e}gles de Feynman en main, nous avons les outils pour
d\'{e}terminer le d\'{e}veloppement diagrammatique de Z. On a, en
effet:
\begin{equation}
\frac{Z}{e^{-\frac{N^2 V_0}{T}} \int_E dM e^{-\frac{N t_2}{2 T} Tr M^2} } = \sum_{\{n_k\}} \left\langle
\prod_k \frac{1}{n_k!}(-\frac{N t_k}{T k} Tr M^k)^{n_k}
\right\rangle_0
\end{equation}

On peut d\'{e}velopper chaque terme de la forme $\left\langle
\prod_k \frac{1}{n_k!}(-\frac{N t_k}{T k} Tr M^k)^{n_k}
\right\rangle_0$ en une somme sur les diagrammes \`{a} $n_k$
vertex \`{a} k lignes externes dont le poids est:
\begin{equation}
\frac{1}{\Omega} \left( \frac{T}{t_2 N}\right)^{n_p} \prod_k
\left(-\frac{N t_k}{T}\right)^{n_k} N^l
\end{equation}
o\`{u} $\Omega$ est le facteur de sym\'{e}trie du diagramme, $n_k$
le nombre de vertex \`{a} k pattes externes, $n_v = \sum_k n_k$ le nombre total
de vertex et l le nombre de boucles du diagramme.

Z s'obtient donc comme la somme de tous les diagrammes form\'{e}s
de $n_k$ vertex \`{a} k pattes, $n_p$ propagateurs et comprenant l
boucles auxquels on a associ\'{e} le poids :
\begin{equation}
\frac{1}{\Omega} N^{n_v-n_p+l} \, T^{n_v-n_p} \, t_2^{n_p}
\prod_k\left(-t_k\right)^{n_k} \label{diag}
\end{equation}

\subsubsection{Lien diagramme-surface.}
Les diagrammes d\'{e}finis plus t\^{o}t peuvent \^{e}tre
interpr\'{e}t\'{e}s de mani\`{e}re g\'{e}om\'{e}trique. En effet,
\`{a} tout vertex \`{a} k pattes, on peut faire correspondre un
polygone \`{a} k c\^{o}t\'{e}s tel que chaque c\^{o}t\'{e}
soit orthogonal \`{a} un propagateur. Par exemple, un vertex \`{a}
trois pattes donne un triangle et un vertex \`{a} 4 lignes
externes donne un carr\'{e}.
\vspace{0.5cm}
\beq
\begin{array}{c}
{\epsfxsize 6cm\epsffile{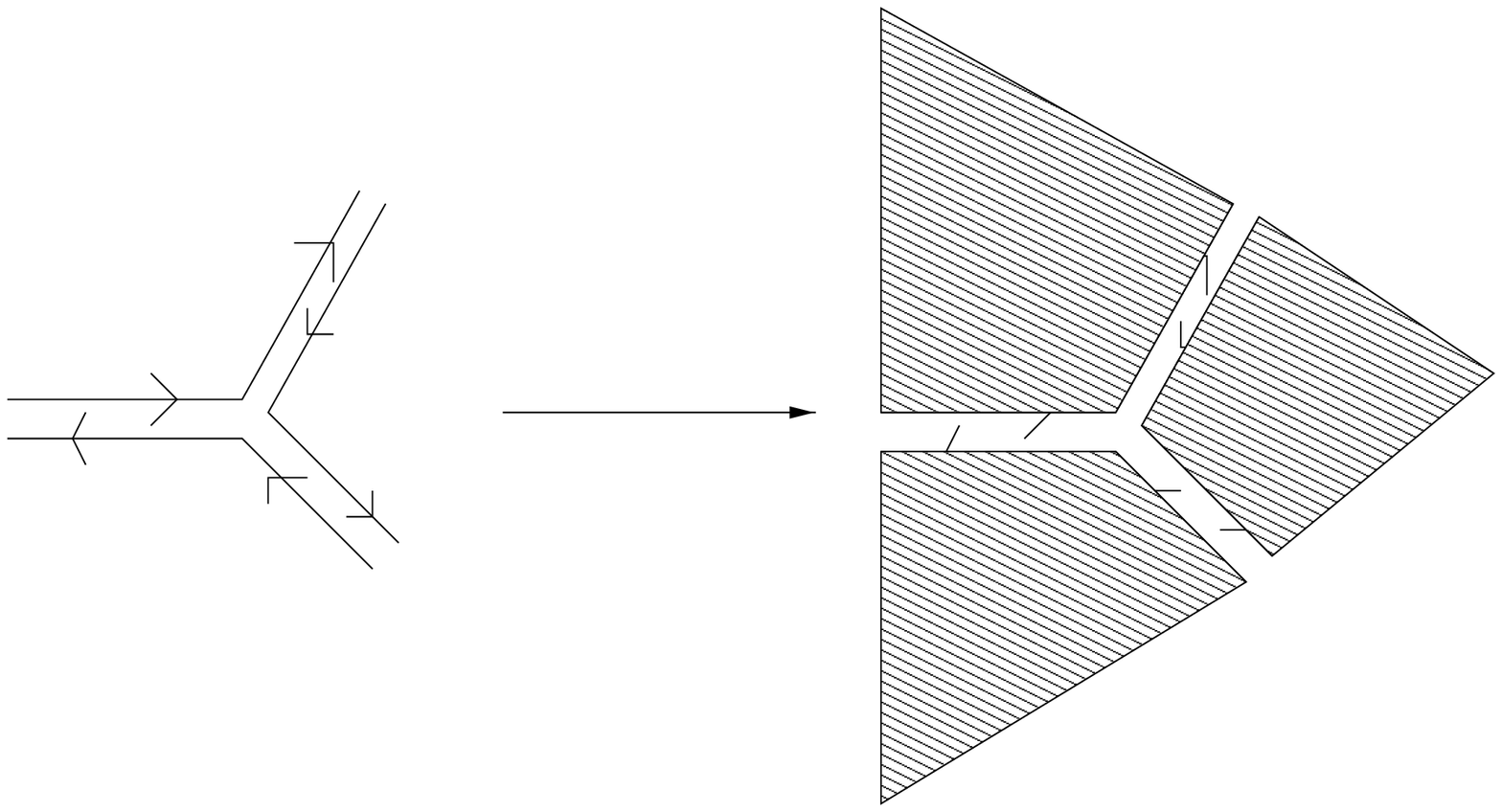}}
\end{array}
\eeq
\vspace{0.5cm}

On obtient ainsi une relation de dualit\'{e} mettant en
correspondance un diagramme avec une surface discretis\'{e}e. Le
d\'{e}veloppement pr\'{e}c\'{e}dent peut alors \^{e}tre vu comme une
sommation sur les surfaces discretis\'{e}es. Ainsi, un diagramme
ne comprenant que des vertex \`{a} trois pattes sera le dual d'une
surface d\'{e}coup\'{e}e selon des triangles.

Peut-on alors avoir une expression directe de Z comme duale de
(\ref{diag})? Ceci n'est possible qu'en s\'{e}parant les
diagrammes non-connexes des diagrammes connexes. Pour ce faire,
introduisons une nouvelle fonction de partition.

Soit $\widetilde{Z} = ln Z$. Le d\'{e}veloppement diagrammatique
de $\widetilde{Z}$ s'obtient en restreignant la sommation obtenue
pour Z sur les diagrammes connexes:
\begin{equation}
\widetilde{Z} = \sum_{diagrammes \,\, connexes} \frac{1}{\Omega}
N^{n_v-n_p+l} \, T^{n_v-n_p} \, t_2^{-n_p}
\prod_k\left(-t_k\right)^{n_k} \label{exprZ}
\end{equation}

$\widetilde{Z}$ peut alors \^{e}tre interpr\'{e}t\'{e} comme la
fonction de partition d'un ensemble de surfaces al\'{e}atoires
tir\'{e}es avec un poids de Boltzmann. Notons les
interpr\'{e}tations des param\`{e}tres d\'{e}finis
pr\'{e}c\'{e}demment:
$t_k$ repr\'{e}sente la fugacit\'{e} des polygones
\`{a} k c\^{o}t\'{e}s;
$t_2$ joue le r\^{o}le d'\'{e}nergie de liaison entre les polygones;
$n_p$ repr\'{e}sente le nombre d'ar\^{e}tes;
$n_v$ est le nombre de polygones constitutifs de la
surface; l est le nombre de sommets;
$\chi = n_v - n_p +l$, exposant de N, repr\'{e}sente la caract\'{e}ristique d'Euler Poincar\'{e} de la surface.

\subsubsection{D\'{e}veloppement topologique en $1/N^2$.}

Le d\'{e}veloppement en puissances de N effectu\'{e} plus haut peut
s'interpr\'{e}ter de mani\`{e}re g\'{e}om\'{e}trique.
Consid\'{e}rons en premier lieu le d\'{e}veloppement
diagrammatique. $\chi = n_v - n_p +l$, l'exposant de N, est en
fait une propri\'{e}t\'{e} de la surface sur laquelle peut
\^{e}tre dessin\'{e} le diagramme consid\'{e}r\'{e} pour ne
pas avoir de lignes qui se croisent. Par exemple, les diagrammes
correspondants \`{a} $\chi=2$ peuvent \^{e}tre trac\'{e}s sur un
plan, $\chi=0$ sur un tore etc... De mani\`{e}re g\'{e}n\'{e}rale,
si l'on note $\chi = 2 - 2 h$, h est le genre d'une telle surface,
c'est \`{a} dire, son nombre de trous. Ainsi, un diagramme de
caract\'{e}ristique $\chi = 2 - 2 h$ ne peut \^{e}tre
dessin\'{e} que sur une surface avec au moins $h$ trous.
\vspace{0.5cm}

De m\^{e}me, dans la description en termes de surfaces
discretis\'{e}es, le d\'{e}veloppement de $\widetilde{Z}$ peut
\^{e}tre vu comme un d\'{e}veloppement topologique. Ainsi:
\begin{equation}
\widetilde{Z} = \sum_{h=0}^{+\infty} N^{2-2h} \widetilde{Z}_h
\end{equation}

o\`{u} $\widetilde{Z}_h$ est la fonction de partition restreinte
aux surfaces polygonales de genre h. On a ainsi un
d\'{e}veloppement topologique selon le genre des surfaces, qui
donne acc\`{e}s aux fonctions $\widetilde{Z}_h$ de genre h
fix\'{e}. Notons que la limite N grand s\'{e}lectionne les
surfaces de genre 0. On l'appelle limite planaire ou limite
sph\'{e}rique.
\beq
\td{Z} = \begin{array}{l}
{\epsfxsize 10cm\epsffile{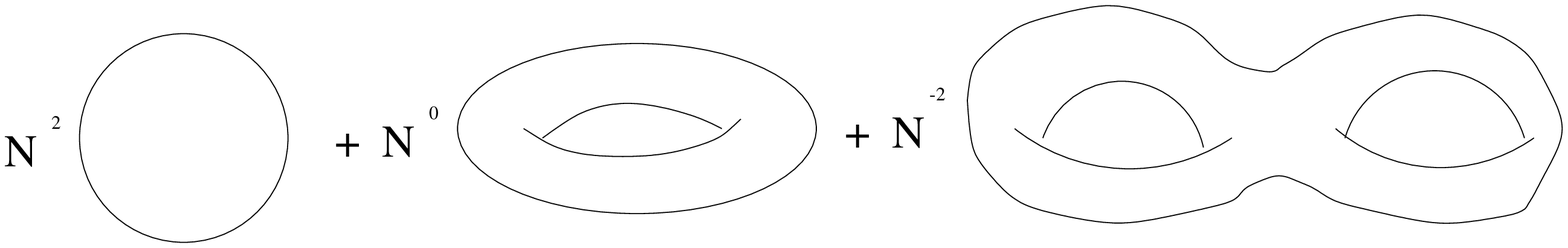}}
\end{array}
+ O(N^{-4})
\eeq

\subsubsection{Interpr\'{e}tation de $\left\langle TrM^n
\right\rangle$.}

Remarquons que le calcul de Z ne fait intervenir que des surfaces
ferm\'{e}es duales de diagrammes eux aussi ferm\'{e}s. Comment
tenir compte des surfaces ouvertes, i.e. avec bords?

Dans ce paragraphe, nous allons introduire de telles surfaces en
nous int\'{e}ressant aux fonctions g\'{e}n\'{e}ratrices :
\begin{equation}
T_n = \left\langle TrM^n \right\rangle = \frac{1}{Z} \int dM \,
\left( Tr M^n \right) e^{-N Tr V(M)}
\end{equation}

Du fait de la pr\'{e}sence du facteur $Tr M^n$, chaque surface
intervenant dans le calcul de $T_n$ contient au moins un polygone
\`{a} n cot\'{e}s. Le reste du diagramme est donc une surface
\`{a} laquelle on a enlev\'{e} un tel polygone. C'est donc une
surface avec un bord de longueur n, et l'on a la relation :
\begin{equation}
\frac{1}{n} T_n = \sum_{\hbox{surface de bord n}} \frac{1}{\Omega}
N^{\chi} \, T^{n_k-n_p} \, t_2^{-n_p}
\prod_k\left(-t_k\right)^{n_k} \label{unbord}
\end{equation}

Que devient le d\'{e}veloppement topologique d\'{e}fini pour les
surfaces sans bords? Il est toujours valable en tenant compte de
la modification dans la d\'{e}finition de la caract\'{e}ristique
d'Euler Poincar\'{e} $\chi$ d'une surface avec un bord : $\chi = 1-2h$.

On peut alors \'{e}crire le d\'{e}veloppement topologique :
\begin{equation}
T_n = \sum_h N^{1-2h} T_{n [h]} = \begin{array}{r}
{\epsfysize 1cm\epsffile{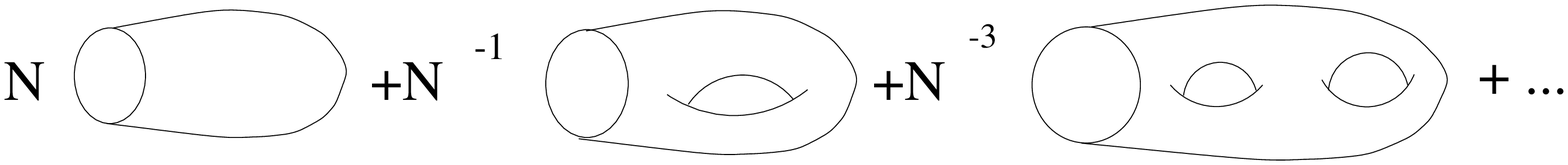}}
\end{array}
\end{equation}

o\`{u} $T_{n [h]}$ est la fonction de partition partielle des
surfaces de genre h ayant un bord de longueur n.

Il sera tr\`{e}s utile pour la suite de d\'{e}finir la
r\'{e}solvante qui jouera le r\^{o}le de fonction de partition
grand canonique :
\begin{equation}
\omega(z) = \frac{1}{N} \sum_{n=0}^{+\infty} \frac{T_n}{z^{n+1}} =
\frac{1}{N} \sum_{n=0}^{+\infty}\left\langle \frac{Tr
M^n}{z^{n+1}} \right\rangle = \frac{1}{N} \left\langle Tr
\frac{1}{z-M} \right\rangle
\end{equation}
\vspace{0.5cm}

On peut g\'{e}n\'{e}raliser la notion de surface avec bord en
consid\'{e}rant les surfaces avec plusieurs bords. Pour ce faire,
introduisons la fonction g\'{e}n\'{e}ratrice des surfaces ayant m
bords de longueurs $l_1,...,l_m$ d\'{e}finie par :
\begin{equation}
T_{l_1,...,l_m} = \left\langle Tr M^{l_1} ... Tr M^{l_m}
\right\rangle_c
\end{equation}

L'\'{e}quation analogue de (\ref{unbord}) s'\'{e}crit alors:
\begin{equation} \label{mbords}
\frac{1}{m! l_1 ... l_m} T_{l_1, ... ,l_m} = \sum_{\hbox{surfaces
de bords $l_1, ... l_m$}} \frac{1}{\Omega} N^{\chi} \, T^{n_k-n_p}
\, t_2^{-n_p} \prod_k\left(-t_k\right)^{n_k}
\end{equation}

o\`{u} l'on red\'{e}fini $\chi = 2-2h-m$.

\subsection{Matrices al\'{e}atoires et th\'{e}orie conforme.}

Dans la partie pr\'{e}c\'{e}dente, nous avons pr\'{e}sent\'{e} le
lien entre matrices al\'{e}atoires et surfaces
discr\'{e}tis\'{e}es. Nous allons maintenant nous int\'{e}resser au
passage \`{a} la limite des surfaces continues et ainsi aborder le
lien avec les exposants critiques et la th\'{e}orie conforme.

\subsubsection{Limite continue.}

Pour obtenir une surface continue, il faut faire tendre le nombre
de pi\`{e}ces formant la surface vers l'infini tout en rendant
leur taille infinit\'{e}simale. Ces comportements singuliers
doivent cependant r\'{e}pondre \`{a} certaines contraintes, telles
que garder l'aire de la surface ( i.e. le nombre de pi\`{e}ces
multipli\'{e} par leur surface ) ainsi que la longueur des bords
finies.

Une telle limite est obtenue lorsque l'un des param\`{e}tres du
potentiel ( respectivement plusieurs param\`{e}tres ) s'approche
d'une valeur critique ( resp. d'un multiplet de valeurs critiques
). On est alors \`{a} un point critique ( resp. multicritique ).

Soit $n_k$ le nombre de polygones \`{a} k cot\'{e}s. En
consid\'{e}rant l'expression de $\widetilde{Z}$ dans
(\ref{exprZ}), on peut \'{e}crire :
\begin{equation}
\left\langle n_k \right\rangle = t_k \frac{\partial ln
\widetilde{Z}}{\partial t_k}
\end{equation}

En un point critique $t_{kc}$ o\`{u} $\left\langle n_k
\right\rangle \rightarrow \infty$, la fonction de $\widetilde{Z}$
est singuli\`{e}re  avec un exposant critique $\alpha$ :
\begin{equation}
\widetilde{Z} = \widetilde{Z}_{reg} + \widetilde{Z}_{sing}
\end{equation}

o\`{u} $\widetilde{Z}_{reg}$ est r\'{e}guli\`{e}re et $\widetilde{Z}_{sing} \propto (t_k -
t_{kc})^{\alpha}$. Le nombre moyen de polygones \`{a} k cot\'{e}s
se comporte alors :
\begin{equation}
\left\langle n_k \right\rangle \propto (t_k - t_{kc})^{\alpha - 1}
\end{equation}

Les exposants $\alpha$ sont universels et d\'{e}crits par la th\'{e}orie des champs conforme.
Leur d\'{e}termination repr\'{e}sente l'un des objectifs principaux
des matrices al\'{e}atoires appliqu\'{e}es aux th\'{e}ories conformes.

\subsubsection{ Action d'Einstein-Polyakov.}

En th\'{e}orie des cordes et gravitation quantique, la fonction de
partition s'\'{e}crit sous la forme (\cite{ZJDFG}):
\begin{equation}
\widetilde{Z} = \sum_{\hbox{surface \cal{S}}} {d\cal{S}} \,\,
e^{-\cal{E}(\cal{S})}
\end{equation}

o\`{u} l'action d'Einstein-Polyakov est d\'{e}finie par :
\begin{equation}
{\cal{E}} = 4 \pi G \chi + \Lambda \, \hbox{aire}\, + \,
\hbox{mati\`{e}re} \label{einstein}
\end{equation}

G \'{e}tant la constante gravitationnelle et $\Lambda$ la
constante cosmologique du mod\`{e}le.

Supposant que les polygones constitutifs de la surface aient des
cot\'{e}s de longueur $\epsilon$, par identification de
(\ref{einstein}) avec (\ref{exprZ}), on obtient :
\begin{equation}
\begin{array}{c}
ln N = - 4 \pi G \\
\Lambda_k \epsilon^2 = ln(-t_k) -\frac{k}{2} ln(t_2)
\end{array}
\end{equation}

Dans ces conditions, l'aire peut s'\'{e}crire ${\cal{A}} =
\left\langle n_t \right\rangle \epsilon^2 $ o\`{u} $n_t = \sum
n_k$ est le nombre total de polygones constituants la surface.
Alors $\left\langle n_t \right\rangle $ doit diverger en
$\frac{1}{\epsilon^2}$. Ceci nous am\`{e}ne \`{a} d\'{e}finir une
constante cosmologique renormalis\'{e}e $\Lambda_{kR}$ telle que :
\begin{equation}
t_k = t_{kc} - \Lambda_{kR} \epsilon^{-\frac{2}{\alpha - 1}}
\end{equation}

et
l'on a l'ordre de grandeur de l'aire moyenne : ${\cal{A}} \propto
\alpha \Lambda_{kR}^{\alpha - 1}$.

\textbf{Note :} En fait, l'exposant critique $\alpha$ introduit
ici, n'est pas bien d\'{e}fini. En effet, un tel exposant, valable
pour tous les genres, n'existe pas. On peut seulement d\'{e}finir
un exposant $\alpha_h$ pour chaque ordre du d\'{e}veloppement
topologique. Cependant, le passage \`{a} la limite continue est
toujours possible en proc\'{e}dant ordre par ordre en
$\frac{1}{N}$.

\subsubsection{Th\'{e}orie conforme et mod\`{e}les minimaux.}
Il existe tr\`{e}s peu de mod\`{e}les que l'on sache r\'{e}soudre compl\`{e}tement (\cite{Gaberdiel:1999mc}): ce sont les mod\`{e}les
minimaux (p,q) dont l'une des caract\'{e}ristiques principales, la charge centrale c, s'\'{e}crit :
\beq
c = 1 - 6 \frac{(p-q)^2}{pq} \,\,\,\, \hbox{avec p et q premiers entre eux.}
\eeq

Les dimensions des champs primaires sont alors les poids conformes de la table de Kac~:
\beq
h_{r,s} = \frac{(rq-sp)^2 -(p-q)^2}{4pq} \,\,\,\, \hbox{avec} \,\,\,0<r<p \,\,\,\, et \,\,\,\, 0<s<q
\eeq

Pour pouvoir comparer les r\'{e}sultats obtenus par les m\'{e}thodes de matrices al\'{e}atoires et de th\'{e}ories conformes,
il faut se ramener \`{a} des quantit\'{e}s calculables dans les deux cas. Ce seront les exposants critiques.
Parmi ceux-ci,  les plus couramment utilis\'{e}s sont les exposants li\'{e}s \`{a} l'aire et \`{a} la topologie, c'est \`{a}
dire le genre h.

On peut, par exemple, consid\'{e}rer le comportement, pour $\cal{A} \rightarrow \infty$, 
de la fonction de partition \`{a} aire et genre fix\'{e}s :
\beq
Z_h({\cal{A}}) \sim_{{\cal{A}} \rightarrow \infty} {\cal{A}}^{\gamma_h -3}
\eeq

Ceci nous permettra de d\'{e}finir la susceptibilit\'{e} de corde $\gamma_{str}$ :
\beq
\gamma_h = 2- (2 - \gamma_{str})(1-h)
\eeq

Dans le cas des mod\`{e}les minimaux, en genre $h=0$, on peut montrer  que l'on a :
\beq
\gamma_{str} = -2 \frac{|p-q|}{p+q-|p-q|}
\end{equation}

Remarquons que, les mod\`{e}les de matrice donnant $\gamma_{str} = -\frac{2}{p+q-1}$, les r\'{e}sultats ne
coincident que pour les mod\`{e}les unitaires.

On peut \'{e}galement s'int\'{e}resser au comportement d'autres op\'{e}rateurs ${\cal{O}}_{r,s}$ vivant sur la surface :
\beq
\moy{{\cal{O}}_{r,s}}({\cal{A}}) \sim {\cal{A}}^{1-\Delta_{r,s}}
\eeq

L'exposant $\Delta_{r,s}$ peut \^{e}tre d\'{e}terminer par les th\'{e}ories conformes si l'op\'{e}rateur
consid\'{e}r\'{e} vit \`{a} l'int\'{e}rieur de la surface. Si c'est un op\'{e}rateur de bord, la m\'{e}thode
des matrices semble \^{e}tre une m\'{e}thode efficace en l'absence de th\'{e}orie g\'{e}n\'{e}rale.. 

\subsubsection{Exemple: cas d'une surface \`{a} deux bords; le
cylindre.}

Dans cette partie, nous allons \'{e}tudier le cas des surfaces
\`{a} deux bords, d\'{e}coup\'{e}es selon des carr\'{e}s, comme exemple d'application.
Notons qu'une surface \`{a} deux bords n'est rien d'autre qu'un
cylindre.

Nous allons donc \'{e}tudier la fonction g\'{e}n\'{e}ratrice des
surfaces \`{a} deux bords, l'un de p\'{e}rim\`{e}tre $L = l
\epsilon$ et l'autre $K = k \epsilon$, o\`{u} $\epsilon$
repr\'{e}sente la taille du cot\'{e} d'un triangle :
\begin{equation}
T_{l,k} = \left\langle Tr M^{l} Tr M^{k} \right\rangle_c
\end{equation}

En utilisant le potentiel quartique sym\'{e}trique:
\begin{equation}
V(M) = \frac{1}{2} M^2 - \frac{g}{4} M^4
\end{equation}

Comme on l'a vu pr\'{e}c\'{e}demment, le nombre moyen de carr\'{e}s de
cot\'{e} $\epsilon$ est :
\begin{equation}
\left\langle n \right\rangle = g \frac{\partial ln
T_{l,k}}{\partial g}
\end{equation}

Ainsi, l'aire moyenne d'un cylindre de bords de p\'{e}rim\`{e}tres $ L =  l \epsilon$ et $L'=l'\epsilon$
sera :
\beq
{\cal{A}} = \epsilon^2 g \frac{\partial ln
T_{l,l'}}{\partial g}
\eeq

Il faut donc d\'{e}terminer les fonctions $T_{2l,2k}(g)$. Pour ce faire, consid\'{e}rons maintenant la fonction \`{a} deux points,
g\'{e}n\'{e}ralisation de la r\'{e}solvante $\omega(z)$ :
\begin{equation}
\omega(x,y) = \left\langle Tr \frac{1}{x-M} Tr \frac{1}{y-M}
\right\rangle_c = \sum_{l=0}^{\infty} \sum_{k=0}^{\infty}
\frac{T_{l,k}}{x^{l+1} y^{k+1}}
\end{equation}

Dans la litt\'{e}rature (\cite{eynard}), on trouve l'expression de $\omega(x,y)$ qui est en fait le noyau de Bergmann (cf le paragraphe qui lui est consacr\'{e}):
\begin{equation}
\omega(x,y) = - \frac{1}{2 (x-y)^2} \left[ 1 - \frac{x y - a^2}{\sqrt{\sigma(x)\sigma(y)}}\right]
\end{equation}

o\`{u} l'on a consid\'{e}r\'{e}, profitant de la sym\'{e}trie du potentiel,
que la densit\'{e} $\rho(x)$ a
pour support le segment $[-a,a]$ (c'est ce que l'on appelle le cas
\`{a} une coupure). On a d\'{e}fini la fonction $\sigma(x) =
x^2-a^2$.

Par \'{e}tude dimensionnelle, on sait que $T_{k,l} = C_{k,l} a^{k+l}$ o\`{u} $C_{k,l}$ est un nombre ind\'{e}pendant de g.
On a ainsi :
\beq
{\cal{A}}(L,K) = \epsilon^2 (l+k) g \frac{\partial ln a }{\partial g}
\eeq

On peut par ailleurs (\cite{eynardthese}) d\'{e}terminer $a^2 = \frac{2}{3g}\left( 1- \sqrt{1-12g} \right)$.
Ce qui nous donne :
\beq
{\cal{A}}(L,K) = \epsilon (L+K) \frac{3 \left[ 4 \frac{1}{\sqrt{1-12g}} - \frac{2}{3} \frac{1-\sqrt{1-12g}}{g} \right]}
{2 g \left( 1- \sqrt{1-12g} \right) }
\eeq

Le passage \`{a} la limite continue n\'{e}cessite de consid\'{e}rer $ \epsilon \rightarrow 0 $ tout en gardant
L, K et ${\cal{A}}$ finis. Pour cela, on doit se placer au voisinage d'un point critique $g \rightarrow g_c = \frac{1}{12}$
avec $g-g_c \sim \epsilon^2$. Soit alors la constante cosmologique renormalis\'{e}e $\Lambda = \frac{1}{L_0^2}$, telle que :
\beq
g = \frac{1}{12}\left( 1- \Lambda \epsilon^2 \right)
\eeq

On obtient alors :
\beq
{\cal{A}}(L,K) = 36 (L+K) L_0 + O(\epsilon)
\eeq

Notons que ce r\'{e}sultat correspond \`{a} celui un cylindre de hauteur $L_0$, ce qui nous renseigne sur la forme de la surface.

\section{Le mod\`{e}le \`{a} deux matrices.}

\subsection{Introduction du mod\`{e}le.}

Ce mod\`{e}le est une extension du mod\`{e}le \`{a} une matrice
dans le sens o\`{u} l'on consid\`{e}re maintenant deux matrices
hermitiennes
$M_1$ et $M_2$ de taille $N \times N $, ainsi que l'int\'{e}grale
:
\begin{equation}
Z = \int dM_1 \, dM_2 \, e^{- N Tr\left(V_1(M_1) + V_2(M_2) - c
M_1 M_2 \right)} \label{int2mat}
\end{equation}
avec $V_1$ et $V_2$ deux potentiels polynomiaux d\'{e}finis de mani\`{e}re
analogue au V pr\'{e}c\'{e}dent.

Tout d'abord, on peut remarquer que l'int\'{e}grale est modifi\'{e}e de fa\c{c}on triviale
par changement de variable affine sur les matrices. Ainsi, on
peut se placer dans le cas o\`{u} les coefficients des termes
quadratiques de $V_1$ et $V_2$ sont tous deux \'{e}gaux \`{a}
$\frac{1}{2}$.

D\`{e}s lors, une interpr\'{e}tation sous forme de diagramme peut
\^{e}tre introduite comme dans le cas du mod\`{e}le \`{a} une
matrice, en prenant soin toutefois de diff\'{e}rencier les deux
matrices. On peut associer deux
couleurs diff\'{e}rentes aux deux matrices, bleu pour $M_1$ et
rouge pour $M_2$. Les diagrammes de Feynman seront alors
l\'{e}g\`{e}rement plus compliqu\'{e}s.

Ils feront ainsi apparaitre trois types de propagateurs
diff\'{e}rents : l'un tout rouge correspondant au lien
$M_2$-$M_2$, l'autre tout bleu pour le lien $M_1$-$M_1$ et enfin
un propagateur mixte avec un cot\'{e} bleu et un cot\'{e} rouge
pour repr\'{e}senter le lien entre $M_1$ et $M_2$. Notons que le
poids de ce troisi\`{e}me propagateur est \'{e}gal \`{a} celui
d'un propagateur unicolore multipli\'{e} par c. Par contre, les
vertex resteront quant \`{a} eux unicolores, aucun terme de la
forme $M_1^{a} M_2^{b}$ avec le produit $ab > 1$ n'apparaissant
dans l'action.

\subsection{Potentiel cubique et mod\`{e}le d'Ising.}
Consid\'{e}rons le cas o\`{u} les deux potentiels sont cubiques.
C'est \`{a} dire, (\ref{int2mat}) s'\'{e}crit :
\begin{equation}
Z = \int dM_1 \, dM_2 \, e^{-N Tr\left( \frac{g_1}{3}M_1^3 +
\frac{g_2}{2} M_2^3 +\frac{1}{2} M_1^2 +\frac{1}{2} M_2^2 -c M_1
M_2 \right)}
\end{equation}

Les diagrammes de Feynman sont alors constitu\'{e}s de deux vertex
\`{a} trois pattes, l'un bleu et l'autre rouge, et des trois propagateurs d\'{e}finis
pr\'{e}c\'{e}demment :
\beq
\begin{array}{c}
{\epsfxsize 15cm\epsffile{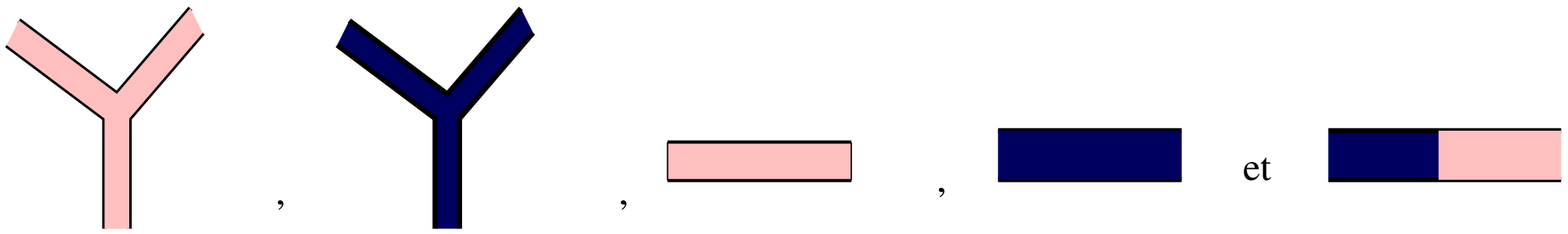}}
\end{array}
\eeq

Le passage aux surfaces duales ne fait plus alors apparaitre de
simples surfaces triangul\'{e}es, mais des surfaces compos\'{e}es
de deux types de triangles : les uns rouges et les autres bleus.
On peut repr\'{e}senter ces diff\'{e}rentes couleurs en attribuant
par exemple un spin +1 aux triangle rouges et un spin -1 aux
triangles bleus. On identifie alors le mod\`{e}le \`{a} deux
matrices \`{a} l'\'{e}tude de surfaces triangul\'{e}es sur
lesquels sont dispos\'{e}s des spins + et -.

\beq
\begin{array}{c}
{\epsfxsize 6cm\epsffile{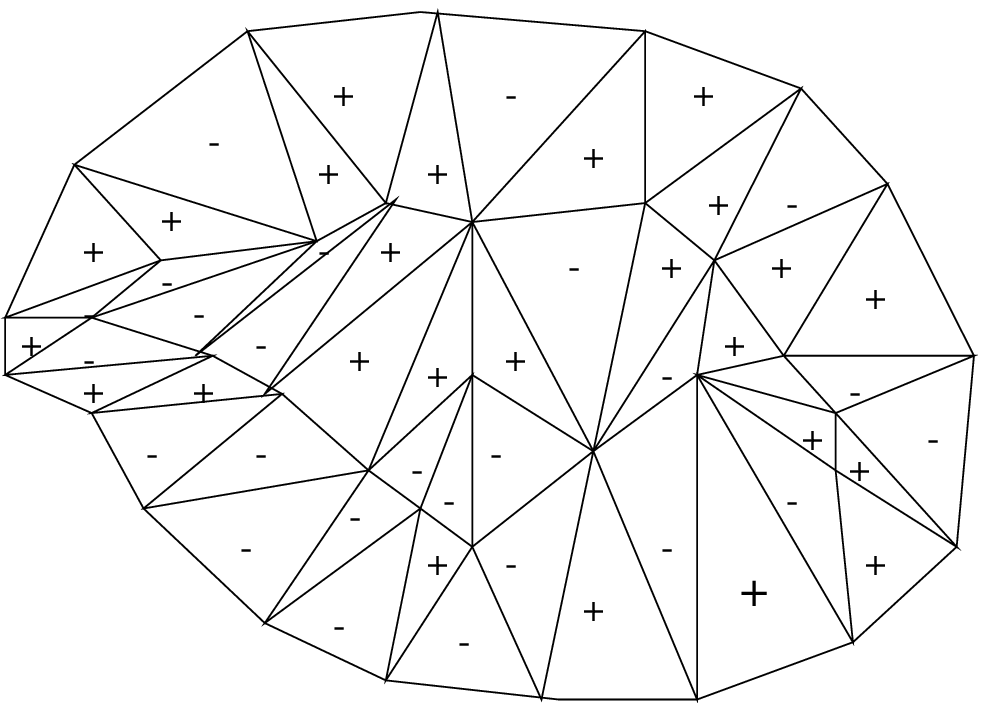}}
\end{array}
\eeq

 On note ainsi
l'analogie avec le mod\`{e}le d'Ising. Il reste \`{a} d\'{e}finir
le lien entre les param\`{e}tres de ces deux mod\`{e}les (\cite{Kazakov},\cite{KazakovIsing}, \cite{DKK}, \cite{KazMar}).

Pour cela, il suffit d'identifier le poids associ\'{e} \`{a} un
diagramme dans chacun des deux mod\`{e}les.
\begin{itemize}
    \item Pour le mod\`{e}le de matrices: un diagramme \`{a}
    $n_\sigma$ triangles de spin $\sigma$, $n_{+-}$ propagateurs
    liant deux triangles de spins diff\'{e}rents et $n_p = \frac{3}{2} (n_+ + n_-)$ propagateurs en tout aura pour poids :
    \begin{equation}
    N^{\chi} \frac{1}{\Omega} g_1^{n_+} g_2^{n_-} c^{n_{+-}} (1-c^2)^{-n_p}
    \end{equation}
    \item Pour un mod\`{e}le d'Ising avec une constante
    cosmologique g, une constante gravitationnelle N, un nombre total de triangle n, un champ
    magn\'{e}tique h et une temp\'{e}rature J :
    \begin{equation}
    N^{\chi} \frac{1}{\Omega} g^n e^{-J \sum \sigma_i \sigma j} e^{-h \sum
    \sigma_i}
    \end{equation}
\end{itemize}
En remarquant que $2 n_{+-} = n_p - \sum \sigma_i \sigma_j $, on
peut identifier:
\begin{equation}
\begin{array}{c}
g = \sqrt{g_1 g_2}\left( \frac{\sqrt{c}}{1-c^2}
\right)^{\frac{3}{2}} \\
e^{-2h} = \frac{g_1}{g_2} \\
E^{2J} =  c
\end{array}
\end{equation}

\subsection{M\'{e}thode des \'{e}quations de boucles et
applications.}
Dans cette partie, nous allons nous int\'{e}resser \`{a} une m\'{e}thode de calcul permettant en partie de r\'{e}soudre le mod\`{e}le 
\`{a} 2 matrices : la m\'{e}thode des boucles (\cite{staudacher}, \cite{eynardchain}, \cite{ZJDFG}, \cite{Akemann}), qui n'est autre que la m\'{e}thode
de Schwinger-Dyson dans le cadre des matrices al\'{e}atoires. Dans un premier temps, nous aborderons cette m\'{e}thode d'un point de vue th\'{e}orique
avant de l'appliquer au calcul de la fonction $\moy{\Tr{\frac{1}{x-M_1}\frac{1}{y-M_2}} \Tr{\frac{1}{x'-M_1}\frac{1}{y'-M_2}}}_c$
que j'ai effectu\'{e} au cours de ce stage.

\subsection{M\'{e}thode des \'{e}quations de boucles.}
Consid\'{e}rons deux potentiels polynomiaux $V_1$ et $V_2$ d\'{e}finis par :
\beq
V_1(x) = \sum_{k=0}^{d_1} \frac{g_{k+1}}{k+1} x^{k+1} \,\,\,\, \hbox{et} \,\,\,\, V_2(y) = \sum_{k=0}^{d_2} \frac{\td{g}_{k+1}}{k+1} y^{k+1}
\eeq

La m\'{e}thode des \'{e}quations de boucles consiste simplement \`{a} effectuer un changement de variable infinit\'{e}simal
de la forme :
 \beq
 M_i \rightarrow \td{M}_i = M_i + \epsilon f(M_1,M_2) \,\,\,\, \hbox{pour} \,\,\,\, i=1,2
 \eeq
 dans les int\'{e}grales de matrice. {\bf Notons que $f(M_1,M_2)$ doit \^{e}tre hermitienne pour que 
$\td{M_i}$ le soit aussi}. On va simplement \'{e}crire l'invariance \`{a} l'ordre 1 en $\epsilon$ 
 de la fonction de partition Z ( on consid\`{e}re ici le cas $i = 1$, l'autre cas \'{e}tant totalement analogue) :
 {\footnotesize
 \bea
 Z & = & \int dM_1\, dM_2\, e^{- N Tr\left(V_1(M_1) + V_2(M_2) -
M_1 M_2 \right)}\cr
& = & \int d\td{M}_1\, dM_2\, e^{- N Tr\left(V_1(\td{M}_1) + V_2(M_2) -
\td{M}_1 M_2 \right)}\cr
& = & \int dM_1\, dM_2\, (1 + \epsilon J(M_1,M_2)) e^{- N Tr\left(V_1(\td{M}_1) + V_2(M_2) -
\td{M}_1 M_2 \right)} (1 - \epsilon K(M_1,M_2)) + O(\epsilon^2)
\eea}

o\`{u} $J(M_1,M_2)$ et $K(M_1,M_2)$ correspondent \`{a} l'ordre 1 en $\epsilon$ respectivement du jacobien du changement de variable
et de la variation de l'action. C'est \`{a} dire :
\bea
\det \frac{\partial \td{M}_1}{ \partial M_1} & = & 1 + \epsilon J(M_1,M_2) + O(\epsilon^2) \cr
N Tr\left(V_1(\td{M}_1) + V_2(M_2) -
\td{M}_1 M_2 \right) & = &N Tr\left(V_1(M_1) + V_2(M_2) -
M_1 M_2\right) \cr
& & + \epsilon K(M_1,M_2) + O(\epsilon^2)\cr
\eea

On obtient ainsi l'\'{e}quation de boucle g\'{e}n\'{e}rique :
\beq
\moy{J(M_1,M_2)}  = \moy{K(M_1,M_2)}
\eeq

\subsubsection{D\'{e}termination de $J(M_1,M_2)$ et de $K(M_1,M_2)$.}
\begin{itemize}
\item
La variation de l'action par ce changement de variable donne ais\'{e}ment l'expression~:
\beq
K(M_1,M_2) = N \left\{ Tr \left[ V_1'(M_1) f(M_1,M_2) - M_2 f(M_1,M_2) \right] \right\}
\eeq

\item
La d\'{e}termination de $J(M_1,M_2)$ est quant \`{a} elle moins \'{e}vidente, mais peut heureusement \^{e}tre
r\'{e}sum\'{e}e par deux r\`{e}gles simples, connues sous le noms de r\`{e}gles "split" et "merge" (\cite{courseynard}).
Celles-ci correspondent en fait aux deux types de changement de variable effectu\'{e}s en pratique.
\vspace{0.5cm}

{\bf{R\`{e}gle Split.}}

Le premier type de changement de variable usuel correspond \`{a}
$f(M_1,M_2) = A \frac{1}{x-M_1} B$ o\`{u} A et B sont des
fonctions de $M_1$ et $M_2$. Alors la correction issue du Jacobien
est :
\bea
J(M_1,M_2) & = & Tr\left(A \frac{1}{x-M_1}\right) Tr\left(
\frac{1}{x-M_1} B\right) \cr
& & + \,\, \hbox{contributions venant de
$A(M_1)$ et $B(M_1)$.}
\eea
Ainsi, chaque fois que l'on rencontre un terme $\frac{1}{x-M_1}$
en dehors d'une trace, on "coupe" l'expression en deux traces en
introduisant un facteur $\frac{1}{x-M_1}$ dans chaque trace.
\vspace{0.5cm}

{\bf{R\`{e}gle merge.}}

On rencontre \'{e}galement des changements de variable du type

$f(M_1,M_2) = A Tr \left(\frac{1}{x-M_1} B \right)$. On a alors :
\bea
J(M_1,M_2) & = & Tr\left(A \frac{1}{x-M_1} B \frac{1}{x-M_1} \right) \cr
& & + \,\, \hbox{contributions venant de $A(M_1)$ et $B(M_1)$.}
\eea

C'est \`{a} dire que, chaque fois que l'on rencontre un terme
$\frac{1}{x-M_1}$ dans une trace, on regroupe toute l'expression
\`{a} l'int\'{e}rieur d'une m\^{e}me trace en rempla\c{c}ant "Tr" par
un duplicata du facteur $\frac{1}{x-M_1}$.

\end{itemize}

\subsubsection{Quelques d\'{e}finitions.}
Les \'{e}quations de boucles faisant intervenir de nombreuses expressions fastidieuses \`{a} \'{e}crire,
il est n\'{e}cessaire d'introduire quelques notations pour que ce rapport soit lisible. Nous consid\'{e}rerons
donc les fonctions suivantes :
{\small{
\beq
W(x) = \frac{1}{N} \moy{ \Tr \frac{1}{x-M_1}} \,\,\,\, \hbox{et} \,\,\,\, \td{W}(y) = \frac{1}{N} \moy{ \Tr \frac{1}{y-M_2}}
\eeq
\beq
W(x,y) = \frac{1}{N} \moy{\Tr{\frac{1}{x-M_1}\frac{1}{y-M_2}}} \,\,\,\, \hbox{et} \,\,\,\, P(x,y) = \frac{1}{N} \moy{\Tr{\frac{V_1'(x)-V_1'(M_1)}{x-M_1}\frac{V_2'(y)-V_2'(M_2)}{y-M_2}}}
\eeq
\beq
U(x,y) = \frac{1}{N} \moy{\Tr{\frac{1}{x-M_1}\frac{V_2'(y)-V_2'(M_2)}{y-M_2}}}
\,\,\,\, \hbox{et} \,\,\,\,
\td{U}(x,y) = \frac{1}{N} \moy{\Tr{\frac{V_1'(x)-V_1'(M_1)}{x-M_1}\frac{1}{y-M_2}}}
\eeq
\beq
W(x,y;x') = \frac{1}{N^2} \moy{\Tr{\frac{1}{x-M_1}\frac{1}{y-M_2}} \Tr{\frac{1}{x'-M_1}}}
\eeq
\beq
U(x,y;x') = \frac{1}{N^2} \moy{\Tr{\frac{1}{x-M_1}\frac{V_2'(y) - V_2'(M_2)}{y-M_2}} \Tr{\frac{1}{x'-M_1}}}
\eeq
\beq
\td{L}(x,y;y') = \frac{1}{N^2} \moy{\Tr{\frac{1}{x-M_1}\frac{1}{y-M_2}} \Tr{\frac{1}{y'-M_2} V_1'(M_1)}}
\eeq
\beq
P(x,y;x',y') = \frac{1}{N^2} \moy{\Tr{\frac{V_1'(x)-V_1'(M_1)}{x-M_1}\frac{V_2'(y)-V_2'(M_2)}{y-M_2}} \Tr{\frac{1}{x'-M_1}\frac{1}{y'-M_2}}}
\eeq
\bea
H(x,y,x',y') & = & \frac{1}{2N}  \moy{\Tr{\frac{1}{x-M_1}\frac{1}{y-M_2}\frac{1}{x'-M_1}\frac{1}{y'-M_2}}} \cr
& & + \frac{1}{2N} \moy{\Tr{\frac{1}{x-M_1}\frac{1}{y'-M_2}\frac{1}{x'-M_1}\frac{1}{y-M_2}}} 
\eea
\bea
F(x,y,x',y') & = & \frac{1}{2N}  \moy{\Tr{\frac{V_1'(x)-V_1'(M_1)}{x-M_1}\frac{1}{y-M_2}\frac{1}{x'-M_1}\frac{1}{y'-M_2}}} \cr
& &  + \frac{1}{2N} \moy{\Tr{\frac{V_1'(x)-V_1'(M_1)}{x-M_1}\frac{1}{y'-M_2}\frac{1}{x'-M_1}\frac{1}{y-M_2}}} 
\eea
\beq
W(x,y;x',y') = \frac{1}{N^2} \moy{\Tr{\frac{1}{x-M_1}\frac{1}{y-M_2}} \Tr{\frac{1}{x'-M_1}\frac{1}{y'-M_2}}}
\eeq
\beq
\td{U}(x,y;x',y') = \frac{1}{N^2} \moy{\Tr{\frac{V_1'(x)-V_1'(M_1)}{x-M_1}\frac{1}{y-M_2}} \Tr{\frac{1}{x'-M_1}\frac{1}{y'-M_2}}}
\eeq}}

Nous aurons \'{e}galement besoin de deux fonctions fondamentales :
\beq
X(y) = V_2'(y) - \td{W}(y) \,\,\,\, \hbox{et} \,\,\,\, Y(x) = V_1'(x) - W(x)
\eeq

\subsubsection{L'\'{e}quation de boucle ma\^\i tresse.}
Parmis les nombreuses \'{e}quations de boucles que l'on peut former,
l'une a une signification particuli\`{e}re et on l'appelle "master
loop equation" (\cite{staudacher}, \cite{eynardchain}).

Consid\'{e}rons le changement de variable :
\begin{equation}
M_2 \rightarrow M_2 + \epsilon \frac{1}{x-M_1}
\end{equation}

L'\'{e}quation de boucle associ\'{e}e est alors simplement :
\begin{equation}
x W(x) -1 = \frac{1}{N} \left\langle Tr \frac{1}{x-M_1}
V_2'(M_2)\right\rangle \label{eqboucle1}
\end{equation}

Le changement de variable (hermitien bien s\^{u}r) :
\begin{equation}
M_1 \rightarrow M_1 + \epsilon \left( \frac{1}{x-M_1}
\frac{V_2'(y)-V_2'(M_2)}{y-M_2} + \frac{V_2'(y)-V_2'(M_2)}{y-M_2}
\frac{1}{x-M_1}\right)
\end{equation}

donne, en utilisant (\ref{eqboucle1}), l'\'{e}quation de boucle :
\begin{equation}
(y-Y(x))U(x,y) = V_2'(y) W(x) - P(x,y) - x W(x) +1 -
\frac{1}{N^2}U(x,y;x) \label{eqboucle2}
\end{equation}

$U(x,y)$ \'{e}tant un polyn\^{o}me en y, il n'a pas de singularit\'{e}
pour y fini, et l'\'{e}quation (\ref{eqboucle2}) prise en $y =
Y(x)$ s'\'{e}crit :
\begin{equation}
\left\{
\begin{array}{l}
y = Y(x) \\
( V_2'(y) -x ) ( V_1'(x)-y) - P(x,y) + 1 = \frac{1}{N^2}
U(x,y;x) \\
\end{array}
\right.
\end{equation}

En notant le polyn\^{o}me de degr\'{e}s ($d_1 + 1 $) en x et ($d_2 + 1
$) en y :
\begin{equation}
E(x,y) = ( V_2'(y) -x ) ( V_1'(x)-y) - P(x,y) + 1
\end{equation}

on peut r\'{e}\'{e}crire l'\'{e}quation pr\'{e}c\'{e}dente en :
\begin{equation}
E(x, Y(x)) = \frac{1}{N^2} U(x,Y(x);x) \label{eqmaitresse1}
\end{equation}
qui est appel\'{e}e \'{e}quation de boucle ma\^\i tresse.

Cette \'{e}quation est fondamentale car elle permet de d\'{e}terminer la fonction Y(x) et ainsi, donne
acc\`{e}s \`{a} toutes les fonctions g\'{e}n\'{e}ratrices de surfaces al\'{e}atoires.
\vspace{1cm}

\textbf{Remarques:}
\begin{itemize}
\item En \'{e}changeant les indices $1 \leftrightarrow 2$ dans tout
le paragraphe pr\'{e}c\'{e}dent, on aurait obtenu une \'{e}quation
analogue \`{a} (\ref{eqmaitresse1}):
\begin{equation}
E(X(y), y) = \frac{1}{N^2} \tilde{U}(X(y),y;y)
\label{eqmaitresse2}
\end{equation}

\item A l'ordre 0 en $\epsilon$, les \'{e}quations
(\ref{eqmaitresse1}) et (\ref{eqmaitresse2}) nous donnent :
\begin{equation}
\left\{ \begin{array}{l} E(x,Y(x)) = 0 \\
E(X(y), y) = 0 \\
\end{array}
\right.
\end{equation}

Les z\'{e}ros de E(x,y) respectivement en x et y seront alors
not\'{e}s $X_0(y),X_1(y),...,X_{d_1}(y)$ et
$Y_0(x),Y_1(x),...,Y_{d_2}(x)$. D'o\`{u} :
\begin{equation}
E(x,y) = -g_{d_1+1} \prod_{i=0}^{d_1} (x-X_i(y)) = -\td{g}_{d_2+1}
\prod_{i=0}^{d_2} (y-Y_i(x))
\end{equation}

\item E(x,y) = 0  d\'{e}fini une courbe alg\'{e}brique. On voit
ainsi apparaitre, pour la premi\`{e}re fois le lien entre
mod\`{e}les de matrices et g\'{e}om\'{e}trie alg\'{e}brique.
\end{itemize}

\subsection{Interpr\'{e}tation des traces et op\'{e}rateurs de bord.}
Comme nous l'avons vu plus haut, dans le mod\`{e}le \`{a} une
matrice, les surfaces \`{a} k bords sont repr\'{e}sent\'{e}es par
les termes de la forme $\left\langle \left(Tr
\frac{1}{x-M}\right)^k \right\rangle_c$.

Nous allons maintenant nous int\'{e}resser \`{a} la
g\'{e}n\'{e}ralisation de cette interpr\'{e}tation dans le
mod\`{e}le \`{a} deux matrices. En effet, ce cas est bien plus
complexe : du fait de la pr\'{e}sence de deux matrices, on peut
observer deux types de traces; certaines dites mixtes font
apparaitre \`{a} la fois les matrices $M_1$ et $M_2$ \`{a}
l'int\'{e}rieur d'une m\^{e}me trace, les autres \'{e}tant dites
non mixtes.

Il existe cependant une proc\'{e}dure simple pour interpr\'{e}ter
les termes de la forme :
\begin{equation} \label{termgene}
\left\langle \left( Tr \frac{1}{x-M_1} \right)^{n_{+}} \left( Tr
\frac{1}{y-M_2} \right)^{n_{-}} \prod_{k=1}^{\infty} \left[ Tr
\left( \frac{1}{x-M_1} \frac{1}{y-M_2} \right)^{k} \right]^{n_k}
\right\rangle_c
\end{equation}

Les exposants introduits ici caract\'{e}risent les surfaces
intervenant dans le d\'{e}veloppement topologique. Ainsi, on
pourrait montrer que :
 $n_{+}$ est le nombre de bords enti\`{e}rement
constitu\'{e}s de segments +; $n_{-}$ est le nombre de bords
enti\`{e}rement constitu\'{e}s de segments -; $n_t = n_+ + n_- + \sum_k n_k$ est le
nombre total de bords;$n_{k}$ est le
nombre de bords constitu\'{e}s de la succession de k groupes de
segments + contigus s\'{e}par\'{e}s par k groupes de segments -
contigus.
Par exemple pour $n_+=0$, $n_-=0$ et $n_2=1$:
 \beq
\begin{array}{c}
{\epsfysize 5cm\epsffile{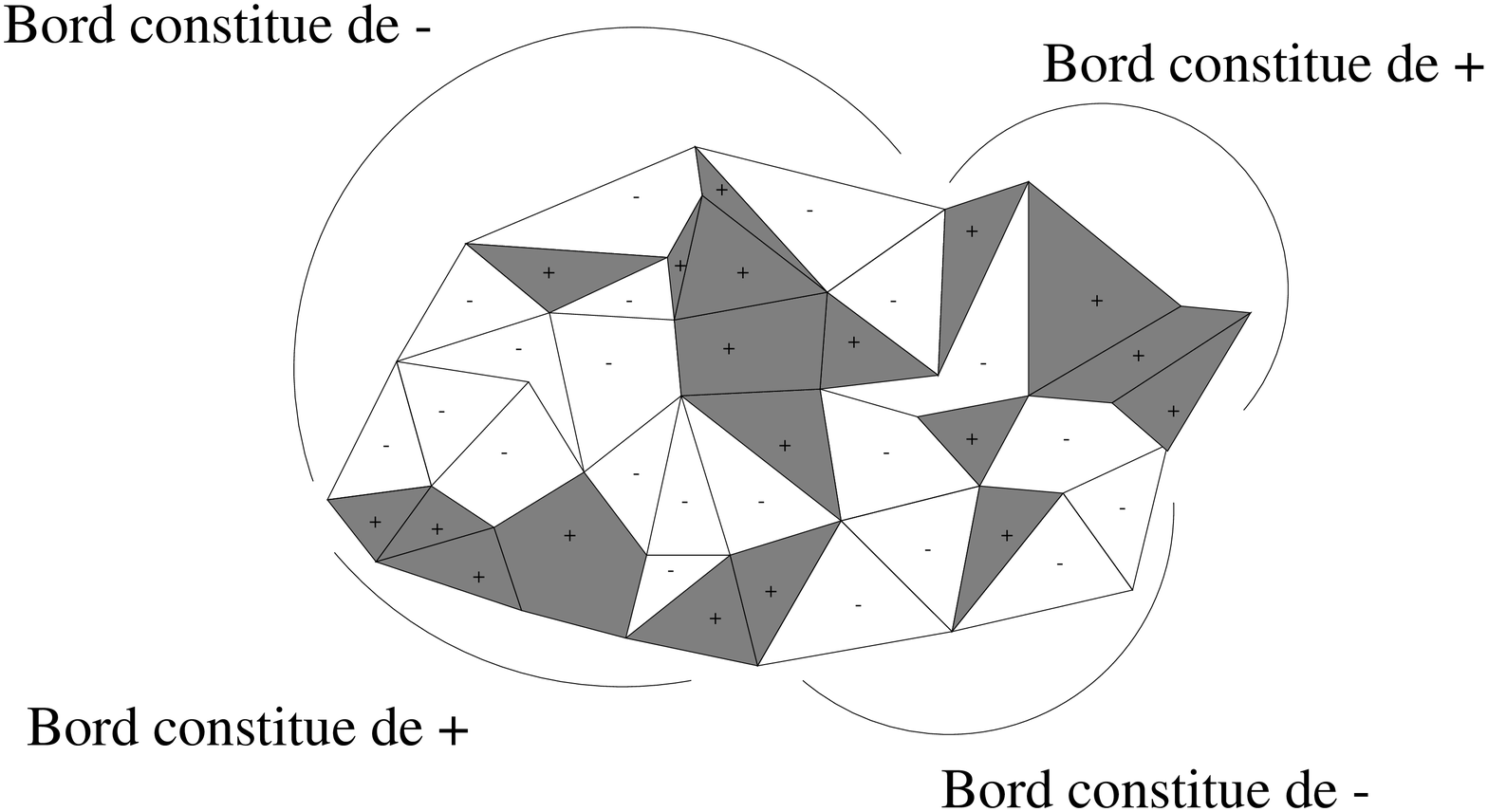}}
\end{array}
\eeq

Ainsi, le nombre de traces indique le nombre de bords tandis
qu'\`{a} l'int\'{e}rieur de chaque trace, le nombre de facteur
$\frac{1}{x-M_1} \frac{1}{y-M_2}$ indique la moiti\'{e} du nombre
d'op\'{e}rateurs de bords, i.e. la moiti\'{e} du nombre de fois
o\`{u} le bord change de couleur.

\subsection{Derniers outils n\'{e}cessaires au calcul: le noyau de Bergmann et les op\'{e}rateurs d'insertion de boucle.}

\subsubsection{Les op\'{e}rateurs d'insertion de boucles.}
On peut se demander comment passer d'une fonction \`{a} une boucle compos\'{e}e de +, W(x), \`{a} une fonction \`{a} deux
boucles +, $\Omega(x;x') = \moy{Tr \frac{1}{x-M_1} Tr \frac{1}{x'-M_1}}$. De mani\`{e}re g\'{e}n\'{e}rale,
on aimerait trouver un op\'{e}rateur qui ins\`{e}re une boucle (de type + ou - ) dans la surface consid\'{e}r\'{e}e.

De tels op\'{e}rateurs existent (\cite{Akemann}, \cite{AmbjAk}). Consid\'{e}rons les d\'{e}rivations par rapport \`{a} $V_1(x)$
et $V_2(y)$ d\'{e}finies formellement par :
\beq
\frac{\partial}{\partial V_1(x)} = - \sum_{k=1}^{\infty} \frac{k}{x^{k+1}} \frac{\partial}{\partial g_k}
\,\,\,\,\,\, \hbox{et} \,\,\,\,\,\,
\frac{\partial}{\partial V_2(y)} = - \sum_{k=1}^{\infty} \frac{k}{y^{k+1}} \frac{\partial}{\partial \td{g}_k}
\eeq

On peut alors remarquer que $\Omega(x;x') = \frac{\partial W(x)}{\partial V_1(x')}$. On pourrait \'{e}galement v\'{e}rifier que la d\'{e}rivation
par rapport \`{a} $V_1(x)$ appliqu\'{e}e sur un terme de la forme (\ref{termgene}) a pour effet
d'augmenter $n_{+}$ d'une unit\'{e}. De m\^{e}me, la d\'{e}rivation per rapport \`{a} $V_2(y)$ augmente
$n_{-}$ de 1. Ainsi, ces op\'{e}rateurs ont pour r\'{e}sultat l'insertion d'un bord (ou boucle) compos\'{e}
respectivement uniquement de + ou uniquement de -. Pour cela on les appelle op\'{e}rateurs d'insertion de boucle.

\subsubsection{Le noyau de Bergmann.}
Consid\'{e}rons \`{a} nouveau la fonction \`{a} deux boucles $\Omega(x;x')$. Elle peut s'\'{e}crire:
\beq
\Omega(x;x') = \frac{\partial W(x)}{\partial V_1(x')} = - \frac{1}{(x-x')^2} - \frac{\partial Y(x)}{\partial V_1(x')}
\eeq

Pour \'{e}tudier le second terme du membre de droite, pla\c{c}ons nous sur la surface de Riemann d\'{e}finie par notre
courbe alg\'{e}brique ${\cal{E}} = \{(x,y) | E(x,y) = 0\}$. Un point p de $\cal{E}$ est alors un couple de complexes
$p = ( {\cal{X}}(p), {\cal{Y}}(p) )$ tels que $E( {\cal{X}}(p), {\cal{Y}}(p) ) = 0 $.

Notons qu'\`{a} chaque x (resp. y) correspondent $d_2+1$ (resp. $d_1 +1$ ) points de ${\cal{E}}$ :
\bea
& {\cal{X}}(p) = x \Leftrightarrow p = p_k(x) \,\,\,\,\,\,\, \hbox{pour k allant de 0 \`{a} $d_2$.} & \cr
& {\cal{Y}}(q) = y \Leftrightarrow q = q_k(x) \,\,\,\,\,\,\, \hbox{pour k allant de 0 \`{a} $d_1$.}
\eea

D\`{e}s lors, on peut d\'{e}finir la fonction sur la surface de Riemann :
\beq
\frac{B(p,p')}{d{\cal X}(p) d{\cal X}(p')} = - \left. \frac{\partial {\cal Y}(p)}{\partial V_1({\cal X}(p'))} \right|_{{\cal X}(p) = cte}
\eeq

Cette fonction a un seul p\^{o}le double en p=p' avec r\'{e}sidu nul sur la surface de Riemann~: c'est
le noyau de Bergmann (\cite{Farkas}, \cite{Fay}), qui s'appelle fonction de Weierstrass en genre 1. 

\section{Application au cas d'un cylindre \`{a} bords bicolores; Calcul de $\moy{\Tr{\frac{1}{x-M_1}\frac{1}{y-M_2}} \Tr{\frac{1}{x'-M_1}\frac{1}{y'-M_2}}}_c$.}
Dans cette partie, je vais pr\'{e}senter ce qui fut l'essentiel de mon travail de recherche au cours de ce stage,
c'est \`{a} dire le calcul de W(x,y;x',y'). Celui-ci permettra par la suite d'avoir des r\'{e}sultats relatifs aux cylindres
dont chaque bords est compos\'{e} de deux segments de couleur diff\'{e}rente.

\beq
\begin{array}{c}
{\epsfysize 3cm\epsffile{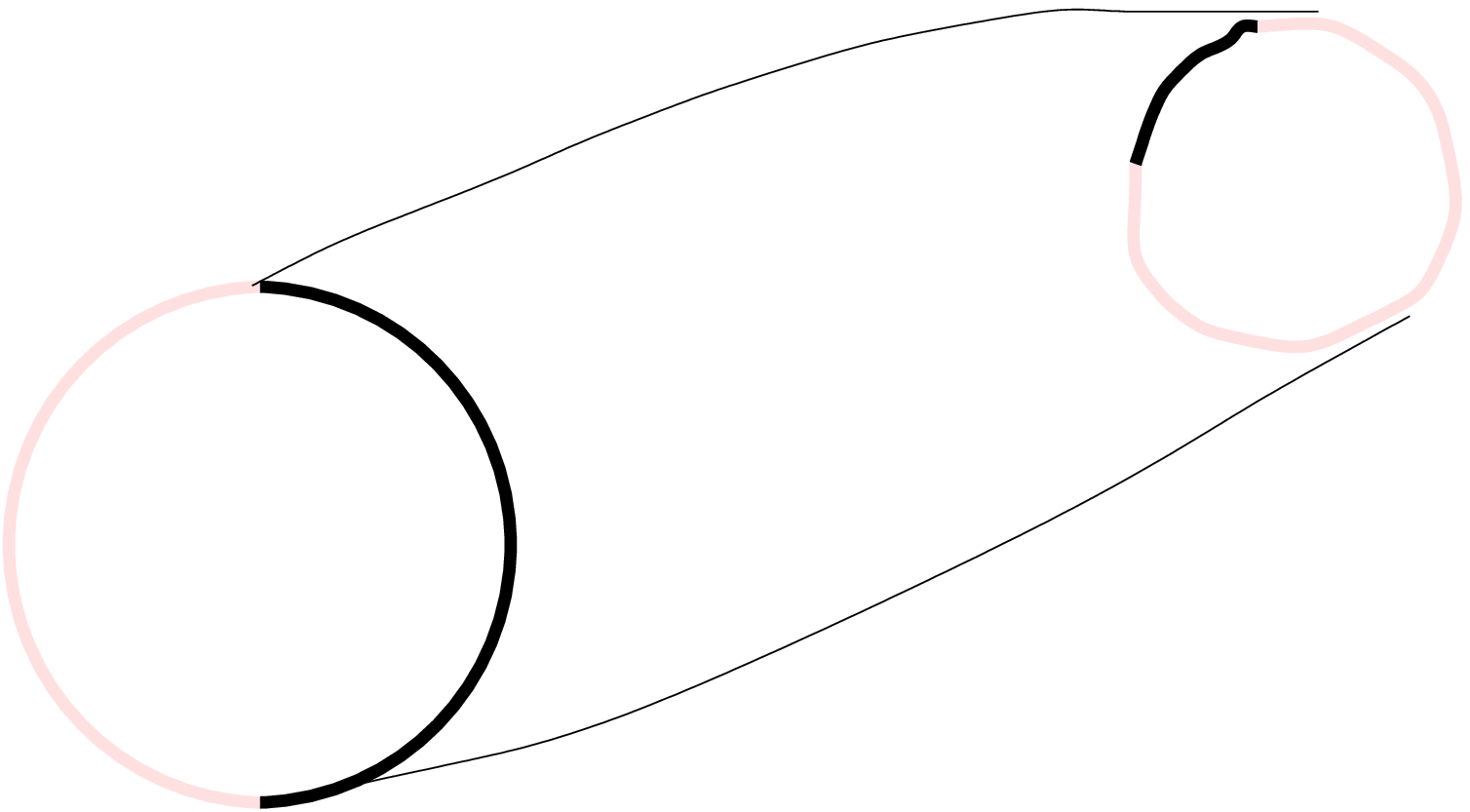}}
\end{array}
\eeq

On pourra ainsi passer \`{a} la limite conforme pour obtenir des informations sur les exposants critiques en pr\'{e}sence
d'op\'{e}rateurs de bords.

Pour effectuer ce calcul, nous allons proc\'{e}der par \'{e}tapes:
\begin{itemize}
\item D\'{e}termination d'un syst\`{e}me d'\'{e}quations de boucles faisant intervenir W(x,y;x',y');
\item R\'{e}solution de ce syst\`{e}me en commenceant par les inconnues dont les propri\'{e}t\'{e}s polynomiales
rendent le calcul plus simple;
\item Simplification de l'expression de W(x,y;x',y') trouv\'{e}e.
\end{itemize}

\subsection{Les \'{e}quations de boucles.}
Dans cette partie, nous allons d\'{e}terminer deux \'{e}quations de boucles dont les seules fonctions inconnues
sont W(x,y;x',y'), $\td{U}(x,y;x',y')$ et P(x,y;x',y'), les autres moments intervenant ayant d\'{e}j\`{a} 
\'{e}t\'{e} calcul\'{e}s dans la litt\'{e}rature.

Notons que ces \'{e}quations ne peuvent pas s'\'{e}crire directement en appliquant la m\'{e}thode des boucles
pr\'{e}sent\'{e}e plus haut. En effet, si usuellement seuls les termes dominants ( de plus haut degr\'{e} en N )
s'annulent gr\^{a}ce \`{a} une \'{e}quation de boucle d'ordre inf\'{e}rieure, ici des termes sous-dominants
disparaissent \'{e}galement car ils interviennent dans cette \'{e}quation d'ordre inf\'{e}rieur.
Ainsi, je pr\'{e}ciserais pour chaque \'{e}quation le changement de variable principal et celui permettant d'\'{e}liminer les termes dominants et certains sous-dominants.

La premi\`{e}re \'{e}quation est obtenue en consid\'{e}rant le changement de variable:
\beq
\delta M_1 = \frac{1}{2} \left[ \frac{1}{x-M_1}\frac{1}{y-M_2} + \frac{1}{y-M_2}\frac{1}{x-M_1} \right] \Tr{\frac{1}{x'-M_1}\frac{1}{y'-M_2}}
\eeq

L'\'{e}quation de boucle associ\'{e}e \`{a} $\partial M_1 = \frac{1}{x-M_1} \frac{1}{y-M_2} + h.c.$, nous permet d'\'{e}crire
\`{a} l'ordre sous dominant :
\bea\label{eqbcl1}
(Y(x)-y) W(x,y;x',y') - \td{U}(x,y';x',y') & = & W(x',y';x) (W(x,y) - 1)\cr
& & + \frac{1}{x-x'} [H(x',y,x',y') - H(x,y,x',y')]\cr
\eea

D'autre part, les changements de variables 
\beq
\delta M_2 = \frac{1}{2} \left[ \frac{V_1'(x)-V_1'(M_1)}{x-M_1}\frac{1}{y-M_2} + \frac{1}{y-M_2}\frac{V_1'(x)-V_1'(M_1)}{x-M_1} \right] \Tr{\frac{1}{x'-M_1}\frac{1}{y'-M_2}}
\eeq

et $\partial M_2 = \frac{V'(x)-V'(M_1)}{x-M_1} \frac{1}{y-M_2} +h.c.$, donnent :
\bea
(X(y)-x) \td{U}(x,y;x',y') - P(x,y;x',y') & = & \td{L}(x',y';y) + \left[ \td{U}(x,y) - V_1'(x) \right] W(x',y';y) \cr
& & + \frac{1}{y-y'} \left[ F(x,y',x',y') - F(x,y,x',y') \right] \cr
\label{eqbcl2}
\eea

\subsection{R\'{e}solution des \'{e}quations.}
Nous allons maintenant r\'{e}soudre ces \'{e}quations en utilisant le fait que P(x,y;x',y') est un polyn\^{o}me en x et y.

Tout d'abord, une autre \'{e}quation de boucle donnant :
\beq
\td{L}(x',y';y) = y W(x',y';y) + H(x',y,x',y')
\eeq

et les r\'{e}sultats $ \td{U}(x,y) - V_1'(x) + y = \frac{E(x,y)}{x-X}$
et $W(x,y) = 1-\frac{E(x,y)}{(x-X)(y-Y)}$ \'{e}tant connus (\cite{eynm2m}' \cite{eynm2mg1}), on peut simplifier (\ref{eqbcl2}) en :
{\footnotesize
\bea \label{eqbcl3}
(X(y)-x) \td{U}(x,y;x',y') - P(x,y;x',y') & = &  \frac{E(x,y)}{x-X} W(x',y';y) \cr
& &+ H(x',y,x',y') + \frac{1}{y-y'} \left[ F(x,y',x',y') - F(x,y,x',y') \right] \cr
\eea}

Les diff\'{e}rents termes du membre de droite peuvent \^{e}tre calcul\'{e}s \`{a} partir
de r\'{e}sultats pr\'{e}existants (\cite{eynm2m}) :
\bea
{\cal{B}} = \frac{E(x,y)}{x-X} W(x',y';y) & = & \frac{E(x,y) E (x',y')}{(x-X)(y'-Y')(x'-X')} \cr
& & * \left[ \sum_{k=1}^{d_1} \frac{\partial X_k(y')}{\partial V_2(y)}\frac{1}{x'-X_k'} - \frac{\partial Y(x')}{\partial V_2(y)} \frac{1}{y'-Y'} \right]
\eea

et 

{\footnotesize{
\bea
& & {\cal{C}} = H(x',y,x',y') + \frac{1}{y-y'} \left[ F(x,y',x',y') - F(x,y,x',y') \right] = \cr
& & \frac{1}{(y-y')(x'-X')(y'-Y')}\left[ \frac{E(x',y)E_x(x',y')-E_x(x',y)E(x',y')}{(x'-X)(y-Y')} \right. \cr
& & \left. + \frac{E(x',y')E_y(x,y')-E_y(x',y')E(x,y')}{(x-X')(x-x')}\right] \cr
& & + \frac{E(x',y)E(x',y')}{(x'-X')(y-Y')(y'-Y')} \left[ \frac{Y_x'}{(x'-X)(y-Y')(y'-Y')} \right. \cr
& & + \left. \frac{X-X'}{(x'-X)^2(x'-X')(y-y')} + \frac{1}{(x-x')(y-y')(x'-X)}\right] \cr
& & + \frac{E(',y')E(x',y')}{(y-y')(x-X')(x'-X')(y'-Y')} \left[ \frac{1}{(x-x')(y'-Y')} - \frac{X_y'}{(x-X')(x'-X')}\right] \cr
& & - \frac{E(x',y')^2}{(x-x')(y-y')(x'-X')^2(y'-Y')^2} - \frac{E(x,y)E(x',y')}{(x-x')(y-y')^2(x-X)(x'-X')(y'-Y')} \cr
& & + \frac{E(x',y)E(x,y')}{(x-x')(y-y')^2(x-X')(x'-X)(y-Y')}
\eea}}

Notons que $\td{U}(x,y;x',y')$ est un polyn\^{o}me en x. Ses p\^{o}les en y ne d\'{e}pendent donc pas de x.
Alors $(X(y)-x)\td{U}(x,y;x',y')$ s'annule pour tout $y = Y_i(x)$. D'autre part, P(x,y;x',y') est un 
polyn\^{o}me en x et y de degr\'{e} $d_2-1$ en y. Il est donc enti\`{e}rement d\'{e}termin\'{e} par sa valeur en
$d_2$ points, par exemple les $Y_i(x)$ pour $i=1..d_2$.

Remarquant que $\left. \frac{E(x,y)}{x-X}\right|_{y=Y_i} = E_x(x,Y_i(x)) $, la formule d'interpolation
prise pour les $y = Y_i(x)$ donne :
\beq
- P(x,y;x',y') = - \sum_{i=1}^{d_2} \frac{P(x,Y_i;x',y') (Y_i-Y)}{(y-Y_i)(y-Y)} \frac{E(x,y)}{E_y(x,Y_i)}
\eeq

Soit, sachant que $ X_y' = - \frac{E_y(X',y')}{E_x(X',y')} $ et $ Y_x' = - \frac{E_x(x',Y')}{E_y(x',Y')} $
{\footnotesize{
\bea
- P(x,y;x',y') &=& \sum_{i=1}^{d_2} \frac{E_x(x,Y_i)}{E_y(x,Y_i)} \frac{E(x,y)E(x',y')(Y_i-Y)}{(y-Y_i)(y-Y)(x'-X')(y'-Y')}\cr
& & * \left[\sum_{k=1}^{d_1} \frac{\partial X_k'}{\partial V_2(Y_i)} \frac{1}{x'-X_k'} - \frac{\partial Y'}{\partial V_2(Y_i)} \frac{1}{y'-Y'}\right] \cr
& & + \frac{(Y_i-Y) E(x,y)}{(x-x')(Y_i-y')(x'-X')(y'-Y')(y-Y_i)(y-Y) E_y(x,Y_i)}\cr
& & * \left[ \frac{E(x',y')E_x(x',Y_i)-E_x(x',y')E(x',Y_i)}{(Y_i-Y')} + \frac{E(x',y')E_y(x,y')-E_y(x',y')E(x,y')}{(x-X')}\right] \cr
& & - \frac{(Y_i-Y)E(x',Y_i)E(x,y)E(x',y')}{(x-x')(x'-X')(Y_i-Y')(y'-Y')(y-Y_i)(y-Y)E_y(x,Y_i)} \cr
& & * \left[ \frac{Y_x'}{(Y_i-Y')(y'-Y')} - \frac{x-X'}{(x-x')(x'-X')(Y_i-y')} + \frac{1}{(x-x')(Y_i-y')}\right] \cr
& & + \frac{(Y_i-Y)E(x,y)E(x,y')E(x',y')}{(Y_i-y')(x-X')(x'-X')(y'-Y')(y-Y_i)(y-Y)E_y(x,Y_i)} \cr
& & * \left[ \frac{1}{(x-x')(y'-Y')} - \frac{X_y'}{(x-X')(x'-X')}\right] \cr
& & - \frac{(Y_i-Y)E(x,y)E(x',y')^2}{(x-x')(Y_i-y')(x'-X')^2(y'-Y')^2(y-Y_i)(y-Y)E_y(x,Y_i)} \cr
& & - \frac{(Y_i-Y)E(x,y)E(x',y')E_x(x,Y_i)}{(x-x')(Y_i-y')^2(x'-X')(y'-Y')(y-Y_i)(y-Y)E_y(x,Y_i)} \cr
& & - \frac{(Y_i-Y)E(x,y)E(x',Y_i)E(x,y')}{(x-x')^2(y-y')^2(x-X')(Y_i-Y')(y-Y_i)(y-Y)E_y(x,Y_i)}
\eea
}}

Nous pouvons alors exprimer W(x,y;x',y') en fonction de P(x,y;x',y') en \'{e}liminant $\td{U}(x,y;x',y')$
dans (\ref{eqbcl1}) et (\ref{eqbcl3}) :
\bea
(Y-y) W(x,y;x',y') & = & \frac{P(x,y;x',y')}{X-x} \cr
& & - \frac{E(x,y)}{x-X} \left( \frac{W(x',y';y)}{x-X} + \frac{W(x',y';x)}{y-Y} \right) \cr
& & + \frac{1}{x-x'} [H(x',y,x',y') - H(x,y,x',y')] \cr
& & + \frac{H(x',y,x',y')}{X-x} \cr
& & + \frac{1}{(y-y')(X-x)} [ F(x,y',x',y') - F(x,y,x',y') ]
\eea

Utilisant les expressions de H(x,y,x',y') et F(x,y,x',y') \'{e}tablies dans \cite{eynm2m} o\'{u} l'on fera tendre x vers x'
et y vers y' ainsi que $ W(x',y';x) = \frac{\partial W(x',y')}{\partial V_1(x)}$ et $ W(x',y';y) = \frac{\partial W(x',y')}{\partial V_2(y)}$,
on obtient finalement l'expression de W(x,y;x',y') :
{\large{
\bea
& (Y-y) W(x,y;x',y') = & \cr
& \frac{E(x,y)E(x',y')}{(x-X)(x'-X')(y'-Y')} \left\{ \sum_{k=1}^{d_1} \frac{1}{x'-X_k'} \left[- \frac{1}{x-X} \frac{\partial X_k'}{\partial V_2(y)} - \frac{1}{y-Y}\frac{\partial X_k'}{\partial V_1(x)} \right] \right. & \cr
& + \sum_{i=1}^{d_2} \frac{E_x(x,Y_i)}{E_y(x,Y_i)} \frac{(Y_i-Y)}{(y-Y_i)(y-Y)} \left[\sum_{k=1}^{d_1} \frac{\partial X_k'}{\partial V_2(Y_i)} \frac{1}{x'-X_k'} - \frac{\partial Y'}{\partial V_2(Y_i)} \frac{1}{y'-Y'}\right] & \cr
& \left. -\frac{1}{y'-Y'} \left[ - \frac{1}{x-X} \frac{\partial Y'}{\partial V_2(y)} - \frac{1}{y-Y}\frac{\partial Y'}{\partial V_1(x)} \right] \right\} & \cr
& + \frac{E(x,y)E(x',y')}{(x-x')(y-y')(x-X)(x'-X')(y'-Y')} \left[ \frac{1}{(x-x')(y-Y)} + \frac{1}{(y-y')(x-X)} \right] & \cr
& - \frac{E(x,y')E(x',y')}{(y-y')(x-X)(x-X')(x'-X')(y'-Y')} \left[\frac{1}{(x-x')(y'-Y')} - \frac{X_y'}{(x-X')(x'-X')} \right] & \cr
& - \frac{E(x',y)E(x',y')}{(x-x')(x-X)(x'-X')(y-Y')(y'-Y')} \left[\frac{1}{(y-y')(x'-X')} - \frac{Y_x'}{(y-Y')(y'-Y')} \right] & \cr
& - \frac{E(x',y)E(x,y')}{(x-x')(y-y')(x-X')(x'-X)(y-Y')} \left[\frac{1}{(y-y')(x-X)} + \frac{1}{(x-x')(y'-Y)} \right] & \cr
& + \frac{E(x',y')^2}{(x-x')(y-y')(x-X)(x'-X')^2(y'-Y')^2} & \cr
& + \frac{E(x',y)E_x(x',y')-E_x(x',y)E(x',y')}{(x-x')(y-y')(y-Y')(x-X)(x'-X')(y'-Y')} 
 - \frac{E(x',y')E_y(x,y')-E_y(x',y')E(x,y')}{(x-x')(y-y')(x-X)(x-X')(x'-X')(y'-Y')} & \cr
& + \frac{1}{x-X} \sum_{i=1}^{d_2}\left\{ \frac{(Y_i-Y) E(x,y)}{(x-x')(Y_i-y')(x'-X')(y'-Y')(y-Y_i)(y-Y) E_y(x,Y_i)} \right. & \cr
& * \left[ \frac{E(x',y')E_x(x',Y_i)-E_x(x',y')E(x',Y_i)}{(Y_i-Y')} + \frac{E(x',y')E_y(x,y')-E_y(x',y')E(x,y')}{(x-X')}\right] & \cr
& - \frac{(Y_i-Y)E(x',Y_i)E(x,y)E(x',y')}{(x-x')(x'-X')(Y_i-Y')(y'-Y')(y-Y_i)(y-Y)E_y(x,Y_i)} & \cr
& * \left[ \frac{Y_x'}{(Y_i-Y')(y'-Y')} - \frac{x-X'}{(x-x')(x'-X')(Y_i-y')} + \frac{1}{(x-x')(Y_i-y')}\right] & \cr
& + \frac{(Y_i-Y)E(x,y)E(x,y')E(x',y')}{(Y_i-y')(x-X')(x'-X')(y'-Y')(y-Y_i)(y-Y)E_y(x,Y_i)} \left[ \frac{1}{(x-x')(y'-Y')} - \frac{X_y'}{(x-X')(x'-X')}\right] & \cr
& - \frac{(Y_i-Y)E(x,y)E(x',y')^2}{(x-x')(Y_i-y')(x'-X')^2(y'-Y')^2(y-Y_i)(y-Y)E_y(x,Y_i)} & \cr
& - \frac{(Y_i-Y)E(x,y)E(x',y')E_x(x,Y_i)}{(x-x')(Y_i-y')^2(x'-X')(y'-Y')(y-Y_i)(y-Y)E_y(x,Y_i)} & \cr
& \left. - \frac{(Y_i-Y)E(x,y)E(x',Y_i)E(x,y')}{(x-x')^2(y-y')^2(x-X')(Y_i-Y')(y-Y_i)(y-Y)E_y(x,Y_i)} \right\} & \cr
\eea
}}

\subsection{Simplification de l'expression de W(x,y;x',y').}
L'expression obtenue plus haut permet un passage \`{a} la limite conforme. Cependant, du fait des nombreuses
sommes sur les $Y_i$, elle n'est pas tout \`{a} fait satisfaisante. On voudrait, en effet, trouver une
expression qui montre que $W(x,y;x',y')$ est une fonction d\'{e}finie sur ${\cal E}$. Pour cela,
il faut que cette derni\`{e}re soit sym\'{e}trique en les $Y_i$. Nous allons donc la simplifier.

Tout d'abord, on peut noter que :
\bea
\sum_{i=1}^{d_2} \frac{Y_i-Y}{(Y_i-y')(y-Y_i)E_y(x,Y_i)} 
& = & \frac{1}{y-y'} \sum_{i=1}^{d_2} \left(\frac{y-Y}{(y-Y_i) E_y(x,Y_i)} - \frac{y'-Y}{(y'-Y_i) E_y(x,Y_i)} \right) \cr
& = & \frac{1}{y-y'} \left(\frac{y-Y}{E(x,y)} - \frac{y'-Y}{E(x,y')} \right)
\eea

Ceci nous permet de calculer trois sommes :
{\large
\beq
\begin{array}{l}
- \sum_{i=1}^{d_2} \frac{(Y_i-Y)E(x,y)E(x',y')^2}{(x-x')(Y_i-y')(x'-X')^2(y'-Y')^2(y-Y_i)(y-Y)E_y(x,Y_i)} = \cr
- \frac{E(x',y')^2}{(x-x')(y-y')(x'-X')^2(y'-Y')^2} + \frac{(y'-Y) E(x',y')^2 E(x,y)}{(x-x')(y-y')(x'-X')^2(y'-Y')^2(y-Y) E(x,y')} \cr
\end{array}
\eeq}

{\large
\beq
\begin{array}{l}
\sum_{i=1}^{d_2} \frac{(Y_i-Y) E(x,y)}{(x-x')(Y_i-y')(x'-X')(y'-Y')(y-Y_i)(y-Y) E_y(x,Y_i)} = \cr
\frac{1}{(x-x')(y-y')(x'-X')(y'-Y')} - \frac{(y'-Y)E(x,y)}{(x-x')(y-y')(x'-X')(y'-Y')(y-Y)E(x,y')}
\end{array}
\eeq}

et

{\large
\beq
\begin{array}{l}
\sum_{i=1}^{d_2} \frac{(Y_i-Y)E(x,y)E(x,y')E(x',y')}{(Y_i-y')(x-X')(x'-X')(y'-Y')(y-Y_i)(y-Y)E_y(x,Y_i)} = \cr
\frac{ E(x,y')E(x',y')}{(y-y')(x-X')(x'-X')(y'-Y')} - \frac{(y'-Y)E(x,y)E(x',y')}{(y-y')(x-X')(x'-X')(y'-Y')(y-Y)}
\end{array}
\eeq}

Pour calculer les termes restants, on doit conna\^\i tre $ A = \sum_{i=0}^{d_2} \frac{(Y_i-Y) E(x',Y_i)}{(Y_i-Y')(y-Y_i)E_y(x,Y_i)} $.
Pour ce faire, on va calculer A de deux mani\`{e}res diff\'{e}rentes.

En \'{e}crivant $Y_i-Y = Y_i-Y'+Y'-Y$, on obtient :
\beq
A  = - 1 + \frac{E(x',y)}{E(x,y)} + (Y'-Y) \sum_{i=0}^{d_2} \frac{E(x',Y_i)}{(Y_i-Y')(y-Y_i)E_y(x,Y_i)}
\eeq

Alors que $Y_i-Y = Y_i-y+y-Y$ donne :
\beq
A  =  - 1 + (y-Y) \sum_{i=0}^{d_2} \frac{E(x',Y_i)}{(Y_i-Y')(y-Y_i)E_y(x,Y_i)}
\eeq

Ceci nous permet d'\'{e}tablir l'expression de A :
\beq
A = - 1 + \frac{y-Y}{y-Y'} \frac{E(x',y)}{E(x,y)}
\eeq
On calcule \'{e}galement :
\beq
B = \sum_{i=0}^{d_2} \frac{(Y_i-Y) E(x',Y_i)}{(Y_i-Y')(y-Y_i)(Y_i-y')E_y(x,Y_i)} = \frac{A(y)-A(y')}{y-y'}
\eeq
\beq
C = \sum_{i=0}^{d_2} \frac{(Y_i-Y) E(x',Y_i)}{(Y_i-Y')^2(y-Y_i)(Y_i-y')E_y(x,Y_i)}
= \frac{\partial B}{\partial Y'}
\eeq 
\beq
D = \sum_{i=0}^{d_2} \frac{(Y_i-Y) E_x(x',Y_i)}{(Y_i-Y')(y-Y_i)(Y_i-y')E_y(x,Y_i)}
= \frac{\partial B}{\partial x'} - Y_x' C
\eeq
et
\beq
E = \sum_{i=0}^{d_2} \frac{(Y_i-Y) E(x',Y_i)}{(Y_i-Y')^2(y-Y_i)E_y(x,Y_i)} 
= \frac{\partial A}{\partial Y'}
\eeq

Le calcul des termes restants est assez diff\'{e}rent. En effet, ces derniers font intervenir des noyaux de
Bergmann et ne d\'{e}pendent pas uniquement de E(x,y).

Tout d'abord, exprimons $ \sum_{k=1}^{d_1} \frac{1}{(x'-X_k') dx } \frac{\partial X_k'}{\partial V_2(y)} - \frac{1}{(y'-Y') dx}\frac{\partial Y'}{\partial V_2(y)}$ en fonction des $Y_i$ 
et non plus des $X_k$ :

\bea
\sum_{k=1}^{d_1} \frac{\partial X_k'}{\partial V_2(y)}\frac{1}{x'-X_k'} & = & - \sum_{k=1}^{d_1} \frac{B(q,q_k')}{dy(q) dy(q_k')} \frac{1}{x(p')-x(q_k')} \cr
& = & - \sum_{k=1}^{d_1}\mathop{\rm Res}_{t \rightarrow q_k'}
\frac{dy(t)}{y(t)-y'} \frac{B(q,t)}{dy(q) dy(t)} \frac{1}{x(p')-x(t)} \cr
& = & \mathop{\rm Res}_{q_0', q, p_i'}
\frac{dy(t)}{y(t)-y'} \frac{B(q,t)}{dy(q) dy(t)} \frac{1}{x(p')-x(t)} \cr
\eea

Ce qui peut s'\'{e}crire :
\bea
\sum_{k=1}^{d_1} \frac{\partial X_k'}{\partial V_2(y)}\frac{1}{(x'-X_k')} & = & \frac{1}{x'-X'} \frac{B(q_0',q)}{dy dy'} - \frac{1}{(x'-X)(y-y')^2} \cr
& & + \sum_{i=0}^{d_2} \frac{B(p_i',q)}{(y'-Y_i') dy dx(p_i')} 
\eea

D'autre part, on sait que :
\beq
\frac{1}{(x-X)}\frac{\partial Y'}{\partial V_2(y)} = \frac{B(p',q)}{(x-X) dx' dy}
\eeq

Et la m\^{e}me m\'{e}thode donne les derniers termes :
\bea
\sum_{k=1}^{d_1} \frac{\partial X_k'}{\partial V_1(x)}\frac{1}{(x'-X_k')} & = & - \frac{1}{x'-X'} \frac{B(q_0',p)}{dx dy'} + \frac{1}{(x'-x)^2(Y-y')} \cr
& & - \sum_{i=0}^{d_2} \frac{B(p_i',p)}{(y'-Y_i') dx dx(p_i')}
\eea 

\bea
\sum_{k=1}^{d_1} \frac{\partial X_k'}{\partial V_2(Y_i)}\frac{1}{(x'-X_k')} & = &  \frac{1}{x'-X'} \frac{B(q_0',p_i)}{dy(p_i)dy'} - \frac{1}{(x'-x)(y'-Y_i)^2} \cr
& & + \sum_{j=0}^{d_2} \frac{B(p_j',p_i)}{(y'-Y_j') dy(p_i) dy(p_j')}
\eea 

\beq
\frac{\partial Y'}{\partial V_2(Y_i)} = \frac{B(p',p_i)}{ dx' dy(p_i)}
\eeq

et

\beq
\frac{1}{(y-Y)}\frac{\partial Y'}{\partial V_1(x)} = - \frac{B(p',p)}{(x-X) dx' dx}
\eeq

Finalement, on obtient pour W(x,y;x',y') l'expression :
{\large
\bea
& (Y-y) W(x,y;x',y') = & \cr
&\frac{E(x,y)E(x',y')}{(x-X)(x'-X')(y'-Y')} \left\{ \frac{B(q',p)}{(y-Y)(x'-X')dx dy'} - \frac{B(q',q)}{(x-X)(x'-X') dy dy'} \right. & \cr
& + \sum_{i=0}^{d_2} \left[ \left( \frac{1}{y-Y} - \frac{1}{y-Y_i} \right) \left( \frac{B(q_0',p_i)}{(x'-X') dy' dx} 
 - \frac{B(p',p_i)}{(y'-Y') dx dx'} + \sum_{j=0}^{d_2} \frac{B(p_j',p_i)}{(y'-Y_j') dx dy(p_j')} \right) \right. & \cr
& \left. \left. - \frac{1}{y'-Y_i'} \left(\frac{B(p_i',q)}{(x-X) dy dx(p_i')} + \frac{B(p_i',p)}{(y-Y) dx dx(p_i')} \right) \right] \right\} & \cr
& + \frac{E(x,y)E(x',y')}{(x-x')(y-y')(x-X)(x'-X')(y'-Y')} \left[ \frac{1}{(x-x')(y'-Y)} + \frac{1}{(y-y')(x'-X)} \right] & \cr
& - \frac{(y-Y)E(x',y)E(x,y')}{(x-x')^2(y-y')^2(x-X')(x'-X)(y-Y')(y'-Y)} + \frac{E(x,y)E(x,y')}{(x-x')^2(y-y')^2(x-X)(x-X')(y-Y)} & \cr
& - \frac{(y'-Y)E(x,y)}{(x-x')(y-y')(x-X)(x'-X')(y'-Y')(y-Y)E(x,y')} \frac{E(x',y')E_y(x,y')-E_y(x',y')E(x,y')}{(x-X')} & \cr
& - \frac{(y'-Y)E(x,y)E(x',y')}{(y-y')(x-X)(x-X')(x'-X')(y'-Y')(y-Y)} \left[ \frac{1}{(x-x')(y'-Y')} - \frac{X_y'}{(x-X')(x'-X')}\right] & \cr
& + \frac{(y'-Y) Y_x' E(x,y)E(x',y')^2}{(x-x')(y-y')(x-X)(x'-X')(y'-Y')^3(y-Y) E(x,y')} & \cr
& - \frac{Y_x' E(x',y')^2}{(x-x')(y-y')(x-X)(x'-X')(y'-Y')(y-Y')^2} & \cr
\eea}

Cette expression ne d\'{e}pendant que de E(x,y) et du noyau de Bergmann, elle est totalement calculable.
Cependant, avant de la d\'{e}terminer \`{a} la limite conforme, il est utile de s'int\'{e}resser d'abord
au cas de W(x,y) dont je n'ai pas encore achev\'{e} le calcul.

\section{Remerciements.}
Pour conclure, je tiens \`{a} remercier Bertrand Eynard qui, au cours de ces quelques mois de stage,
m'a beaucoup appris et guid\'{e} que ce soit par l'interm\'{e}diaires des cours qu'il me donnait ou gr\^{a}ce
aux nombreuses id\'{e}es de probl\`{e}mes qu'il m'a propos\'{e}es.



\begin{thebibliography}{99}




\bibitem{Akemann} G. Akemann, ''Higher genus correlators for the hermitian matrix model with
multiple cuts'', {\em Nucl.Phys.} B482 (1996) 403-430, hep-th/9606004.

\bibitem{AmbjAk} J. Ambjorn, G. Akemann, ''New universal spectral correlators'', {\em J.Phys.} A29 (1996) L555-L560,
cond-mat/9606129.




\bibitem{Antoniadis:1999y} I. Antoniadis, G. Ovarlez, An introduction to perturbative and non-perturbative string theory, hep-th/9906108.

\bibitem{BFSS} T. Banks, W. Fischer, S.H. Shenker, L. Susskind, Phys. Rev. D55 (1997) 5112, hep-th/9610043.


\bibitem{Bilal:1997fy} A. Bilal, M(atrix) theory: A pedagogical introduction, Fortsch. Phys. 47, (1999), 5, hep-th/9710136.


\bibitem{KazakovIsing} D.V.  Boulatov and V.A. Kazakov, ``The
Ising model on a random  planar lattice: the structure of the phase transition
and the exact critical  exponents'', {\em Phys. Lett. B} {\bf 186}, 379 (1987).

\bibitem{BIPZ} E. Brezin, C. Itzykson, G. Parisi, and J. Zuber,
{\em Comm. Math. Phys.} {\bf 59}, 35  (1978).

\bibitem{DKK} J.M. Daul, V. Kazakov, I.K. Kostov, ``Rational
Theories of 2D Gravity from the Two-Matrix Model'', {\em Nucl. Phys.} {\bf
B409}, 311-338 (1993), hep-th/9303093.

\bibitem{Matrixsurf} F. David, ``Planar diagrams, two-dimensional
lattice gravity and surface models'', {\em Nucl. Phys.} {\bf B 257
[FS14]} 45 (1985).

\bibitem{RMTGQ} F. David, Nuc.Phys. B257 (1985) 543, V. Kazakov, Phys. Lett. 150B (1985) 282, J. Ambj\o rn, B. Durhuus, J. Fr\"olich, Nuc. Phys. B259 (1985) 433.

\bibitem{ZJDFG} P. Di Francesco, P. Ginsparg, J. Zinn-Justin,
``2D Gravity and Random Matrices'',
{\em Phys. Rep.} {\bf 254}, 1 (1995).

\bibitem{DVV} R. Dijkgraaf, E. Verlinde, H. Verlinde, Nucl. Phys. B500 (1997) 43, hep-th/9703030.

\bibitem{Dijgrafvafa} R. Dijkgraaf, C. Vafa,
''A Perturbative Window into Non-Perturbative Physics'', hep-th/0208048,
''On Geometry and Matrix Models'', {\em Nucl.Phys.} {\bf B644} (2002) 21-39, hep-th/0207106,
''Matrix Models, Topological Strings, and Supersymmetric Gauge Theories'', {\em Nucl.Phys.} {\bf B644} (2002) 3-20, hep-th/0206255.


\bibitem{courseynard} B. Eynard ``An introduction to random
matrices'', lectures given at Saclay, October 2000, notes available at
http://www-spht.cea.fr/articles/t01/014/.

\bibitem{eynardthese} B. Eynard ``Gravitation quantique bidimensionnelle
et matrices al\'eatoires'',
Th\`ese de doctorat de l'universit\'e Paris 6 (1995).

\bibitem{eynm2m} B. Eynard, ``Large N expansion of the 2-matrix model'',
{\em JHEP} {\bf 01} (2003) 051, hep-th/0210047.

\bibitem{eynm2mg1} B. Eynard, ``Large N expansion of the 2-matrix model,
multicut case'',
preprint SPHT03/106, ccsd-00000521, math-ph/0307052.

\bibitem{eynard} B. Eynard, ``Eigenvalue distribution of large
random matrices, from one matrix to several coupled matrices''
{\em Nucl. Phys. B}{\bf 506}, 633 (1997), cond-mat/9707005.

\bibitem{eynardchain} B. Eynard, ``Correlation functions of
eigenvalues of
multi-matrix models, and the limit of a time dependent matrix'',
{\em J. Phys. A: Math. Gen.} {\bf 31}, 8081 (1998),
cond-mat/9801075.

\bibitem{Farkas} H.M. Farkas, I. Kra, ''Riemann surfaces'' 2nd edition,
Springer Verlag, 1992.

\bibitem{Fay} J.D. Fay, ''Theta functions on Riemann surfaces'',
Springer Verlag, 1973.

\bibitem{Gaberdiel:1999mc} M. R. Gaberdiel, An introduction to conformal field theory, Rept. Prog. Phys. 63 (2000) 607, hep-th/9910156.

\bibitem{GinspargGQ2D}
P. Ginsparg, {\em Matrix models of 2D gravity}
(Trieste Summer School, July  1991).

\bibitem{grossGQ2D} {\em Two dimensional quantum gravity and random surfaces},
edited by D. Gross, T. Piran, and S. Weinberg
(Jerusalem winter school, World Scientific, 1991).

\bibitem{Guhr} T. Guhr, A. Mueller-Groeling, H.A. Weidenmuller,
``Random matrix theories in quantum physics: Common concepts'',
{\em Phys. Rep.} {\bf 299}, 189 (1998).

\bibitem{IKKT} N. Ishibashi, H. Kawai, Y. Kitazawa, A. Tsuchiya, Nucl. Phys. B498 (1997) 467, hep-th/9612115.

\bibitem{Kazakov} V.A. Kazakov, ``Ising model on a dynamical
planar random lattice: exact solution'',
{\em Phys Lett.} {\bf A119}, 140-144 (1986).

\bibitem{KazMar} V.A. Kazakov, A. Marshakov, ''Complex Curve of the
Two Matrix Model and its Tau-function'',
{\em J.Phys.} {\bf A36} (2003) 3107-3136, hep-th/0211236.

\bibitem{KPZ} V.G. Knizhnik, A.M. Polyakov, A.B. Zamolodchikov, Mod. Phys. Lett. A3 (1988) 819.


\bibitem{kunz} H. Kunz, Matrices al\'eatoires en Physique, Cahiers de physique, presses polytechniques et universitaires romandes, 1998.


\bibitem{Mehta} M.L. Mehta, {\em Random Matrices},2nd edition,
(Academic
Press, New York, 1991).


\bibitem{staudacher} M. Staudacher,
`` Combinatorial solution of the 2-matrix model'',
{\em Phys. Lett.} {\bf B305} (1993) 332-338.

\bibitem{thoft} G. 't Hooft, {\em Nuc. Phys.} {\bf B72}, 461 (1974).

\bibitem{Verbaarshot} J.J.M. Verbaarshot, ``Random matrix model approach to chiral symmetry'',
{\em Nucl. Phys. Proc. Suppl.} {\bf 53}, 88 (1997).





\end{thebibliography}
\end{document}